\documentclass{aa}
\usepackage{graphicx}
\usepackage{txfonts}
\usepackage{booktabs}
\usepackage[section]{placeins}           % useful with \FloatBarrier, to keep 
\usepackage{hyperref}
\hypersetup{colorlinks=true, citecolor=blue}

\newcommand{\borus}{\textsc{borus}}
\newcommand{\gro}{GRO J1655-40}
\newcommand{\gx}{GX 339-4}
\newcommand{\laedd}{$\lambda_{\mathrm{Edd}}$}    
\newcommand{\lbol}{$L_{\mathrm{bol}}$}
\newcommand{\ledd}{$L_{\mathrm{Edd}}$}

\newcommand{\mbh}{$M_\mathrm{BH}$}
\newcommand{\msigma}{$M_\mathrm{BH}\text{--}\sigma_\star$}

\newcommand{\nbmc}{$N_\mathrm{BMC}$}
\newcommand{\nustar}{{\it NuSTAR}}
\newcommand{\veldis}{$\sigma_\star$}
\newcommand{\xspec}{\textsc{xspec}}
\newcommand{\xte}{XTE J1550-564}

\begin{document}

%%%%%%%%%%%%%%%%%%%%%%%%%%%%%%%%%%%%%%%%
% if you use custom commands in your title,
% ensure to check your title when submitting!
%%%%%%%%%%%%%%%%%%%%%%%%%%%%%%%%%%%%%%%%
   \title{The black hole--galaxy correlations of the 12\,$\mu$m local AGN sample: a NuSTAR perspective}
   %\subtitle{Subtitle}
%%%%%%%%%%%%%%%%%%%%%%%%%%%%%%%%%%%%%%%%
% Please separate each author with the \and command
%
% Please do not include ORCIDs next to author names.
% Only ORCIDs authenticated by individual authors in EDPS
% editorial system will be taken into account.
% ORCIDs included here will be removed.
%%%%%%%%%%%%%%%%%%%%%%%%%%%%%%%%%%%%%%%%
   \author{Mario Gliozzi\inst{1}\fnmsep\thanks{Corresponding author: mgliozzi@gmu.edu}
        \and Luigi Spinoglio\inst{2}
        \and Juan Antonio Fern\'andez-Ontiveros \inst{3}
        \and Matthew A. Malkan\inst{4}
        \and James K. Williams \inst{1}
        \and Michael Young \inst{1}
        \and Alexandra D. Oxley \inst{1}
        \and Shobita Satyapal \inst{1}
        }
\authorrunning{Gliozzi et al.}
   \institute{Department of Physics and Astronomy, George Mason University, 4400 University Drive, Fairfax, VA 22030, USA\\
              \email{mgliozzi@gmu.edu}
   \and 
   Istituto di Astroﬁsica e Planetologia Spaziali (INAF–IAPS), Via Fosso del Cavaliere 100, I–00133 Roma, Italy
   \and 
   Centro de Estudios de F\'isica del Cosmos de Arag\'on (CEFCA), Plaza San Juan 1, E–44001, Teruel, Spain
   \and
   Department of Physics and Astronomy, UCLA, Los Angeles, CA 90095-1547, USA
   }
   
%   \date{Received ..., 20XX}
% \abstract{}{}{}{}{}
% 5 {} token are mandatory
 
  \abstract
{The M--sigma correlation of inactive galaxies plays a crucial role in estimating the black hole mass (\mbh) in active galactic nuclei (AGN): 1) it is used to calibrate the reverberation mapping (RM) method; 2) the RM is the basis of the single epoch method, the most-used technique to determine \mbh\ of broad-line AGN; and 3) in narrow-line AGN, \mbh\ is often directly inferred from the M--sigma correlation.}
{The main goal of this work is to test whether the M--sigma relation calibrated from inactive galaxies can be reliably applied to a local
AGN population spanning Seyfert~1, Seyfert~2, and hidden broad-line region galaxies, and whether the resulting AGN relation has sufficient
predictive power to be used as an indirect \mbh\ estimator.}
{To investigate this hypothesis, we used the 12\,$\mu$m AGN sample, a local representative and largely unbiased sample. Since a substantial fraction of this sample has good-quality \nustar\ data, we estimated their \mbh\ with the X-ray scaling method, which works for any AGN accreting at moderate or high level.}
{We find that the \msigma\ correlation of the 12\,$\mu$m AGN sample is different from the one obtained with inactive galaxies and has a substantial scatter. Therefore, its predictive power for \mbh\ estimates is fairly limited. We also find that the \mbh--$M_\star$ correlation based on the 12\,$\mu$m AGN sample is consistent with the analogous correlation obtained from a large sample of nearby type 1 AGN and inconsistent with that of inactive galaxies. Finally, we find suggestive evidence that Seyfert 1 and hidden broad-line region galaxies have similar and systematically higher Eddington ratios than Seyfert 2 galaxies. }{}


   \maketitle
\nolinenumbers
%%%%%%%%%%%%%%%%%%%%%%%%%%%%%%%%%%%%%%%%%%
%%%% SECTION 1: INTRODUCTION    
%%%%%%%%%%%%%%%%%%%%%%%%%%%%%%%%%%%%%%%%%%
\section{Introduction}
\label{sec1:Introduction}
%%%%%%%%%%%%%%%%%%%%%%%%%%%%%%%%%%%%%%%%%%
Massive black holes are found at the centers of galaxies spanning a wide range of morphologies, including systems with prominent bulges and some with little or no classical bulges \citep[e.g.,][]{Magorrian1998,Gebhardt2000,Ferrarese2000,Kormendy2013,Peterson2005}. In galaxies with reliable dynamical black hole mass measurements, the black hole mass, \mbh, correlates with several properties of the host-galaxy spheroid, including its stellar velocity dispersion \citep{Gebhardt2000}, bulge luminosity \citep{Kormendy1995}, and bulge stellar mass \citep{Magorrian1998}. Among these relations, the \mbh--\veldis\ correlation is generally the tightest and most widely used \citep[e.g.,][]{Gebhardt2000,Ferrarese2000,Gueltekin2009,Kormendy2013}. 

These empirical correlations connect physical scales that differ by several orders of magnitude and have therefore motivated the idea that the growth of supermassive black holes and the evolution
of their host galaxies are linked.
For example, the radius of the black hole sphere of influence is $R_{\rm infl}=GM_{\rm BH}/\sigma_\star^2$, where $G$ is the gravitational constant and \veldis\ is the stellar velocity dispersion of the bulge. For nearby massive galaxies, $R_{\rm infl}$ is typically only $\sim10$--50 pc, whereas the host spheroid extends over several kiloparsecs. The existence of correlations across such different physical scales has motivated models in which black hole growth and host-galaxy evolution are coupled through a shared gas reservoir that can fuel both star formation and black hole accretion, as well as through galaxy assembly and AGN feedback. In feedback scenarios, radiation, winds, and jets produced by the AGN can heat, ionize, or expel interstellar gas, thereby altering both star formation and subsequent black hole fueling \citep[e.g.,][]{Silk1998,DiMatteo2005,Fabian2012,Heckman2014,King2015}.
Cosmological simulations commonly include AGN feedback in order to reproduce the observed properties of massive galaxies and black hole--galaxy scaling relations \citep[e.g.,][]{Schaye2015,Volonteri2016,McAlpine2018}. The observed correlations do not, however, by themselves establish a unique causal mechanism, and their inferred forms can also depend on galaxy structure, sample selection, and the limited range of black hole masses accessible to dynamical measurements \citep[e.g.,][]{Lauer2007,Graham2008,Graham2012,Kormendy2013,Shankar2016}.

Accurate black hole masses are essential for determining Eddington ratios, constructing black hole mass functions, quantifying black hole growth, and investigating how accretion and feedback depend on host-galaxy properties \citep[e.g.,][]{Kelly2013,Heckman2014}. Direct black hole mass measurements from spatially resolved stellar or gas dynamics require the black hole sphere of influence to be resolved. They are therefore limited primarily to nearby galaxies, many of which have weak or absent nuclear activity. Throughout this paper, we use the term ``inactive galaxy'' to refer to a galaxy without a luminous AGN; it does not imply that the galaxy necessarily lacks star formation. These dynamically measured systems form the principal samples used to establish the standard local black hole--host galaxy relations \citep[e.g.,][]{Gueltekin2009,Kormendy2013}.

For variable AGN with directly visible broad permitted emission lines, hereafter type 1 AGN, \mbh\ can be measured using reverberation mapping (RM; e.g., \citealt{Peterson1993,Peterson2004}). Reverberation mapping measures the time delay, $\tau$, between variations in the ionizing continuum and the response of the broad-line region (BLR), together with a characteristic velocity of the variable line profile, $\Delta V$. The resulting mass is $M_{\rm BH}=f\,c\tau\Delta V^2/G$, where $c$ is the speed of light and $f$ is a virial factor that accounts for the geometry, inclination, and kinematics of the BLR. Because these properties are not generally known for individual sources, a population-averaged virial factor, $\langle f\rangle$, is commonly introduced. Generally, $\langle f\rangle$ is calibrated by requiring RM AGN to follow an adopted \mbh--\veldis\ relation \citep[e.g.,][]{Onken2004,Woo2010,Woo2013}. Consequently, in the local universe, the absolute normalization of the RM mass scale has traditionally depended on the assumption that active and inactive galaxies follow the same underlying \mbh--\veldis\ relation.

RM campaigns require repeated spectroscopic observations over a cadence and duration sufficient to measure the delay between the ionizing continuum and the emission line variations \citep[e.g.,][]{Horne2004}. They can therefore be carried out for only a limited number of suitable broad-line AGN. The empirical correlation between BLR radius and AGN continuum luminosity \citep[e.g.,][]{Laor1998,Wandel1999,Kaspi2005,Bentz2006} extends virial mass estimates to much larger samples through the single-epoch method \citep[e.g.,][]{Vestergaard2002,Vestergaard2006,Shen2024}. In this approach, a single spectrum supplies both a luminosity-based estimate of the BLR radius and a broad-line velocity width. 

Although single-epoch masses have transformed studies of black hole demographics, they inherit the calibration and systematic uncertainties of the RM mass scale and can only be applied when an appropriate broad emission line is directly observable.
This limitation is especially important for type 2 AGN, whose optical spectra do not show directly visible broad permitted emission lines. Type 2 and otherwise obscured AGN constitute a large---and, at Seyfert-like luminosities, often dominant---fraction of the actively accreting black hole population. The precise obscured fraction depends on luminosity, redshift, wavelength of selection, and the adopted classification criterion, but population studies consistently show that obscured systems comprise at least a substantial fraction, and over broad regions of parameter space the majority, of the AGN population \citep[e.g.,][]{Maiolino1995,Hasinger2008,Ueda2014,Buchner2015,Ricci2017a}. Obtaining reliable \mbh\ values and Eddington ratios for this demographic is therefore necessary for a complete census of black hole growth. Without such measurements, comparisons of accretion rates between unobscured and obscured AGN, tests of AGN unification, and estimates of the black hole mass and Eddington-ratio distributions remain systematically incomplete.

Direct dynamical measurements are available for only a small subset of nearby type 2 AGN. An important example is provided by systems with water megamasers tracing nearly Keplerian rotation in spatially resolved circumnuclear disks, which can yield particularly precise \mbh\ measurements \citep[e.g.,][]{Greenhill2003,Kuo2011}. Some type 2 AGN also reveal broad emission lines at infrared wavelengths, where the effects of dust extinction are reduced, or in polarized light. The latter are conventionally referred to as hidden broad-line region (HBLR) AGN. Nevertheless, these methods are applicable to only a minority of type 2 systems. For most of the obscured AGN population, \mbh\ must therefore be inferred using an indirect estimator, most commonly the \mbh--\veldis\ relation calibrated from inactive galaxies.

It is consequently important to address two related but different questions. The first is whether the whole AGN population (including type 1 and type 2 AGN) exhibits a statistically significant correlation between \mbh\ and \veldis. The second, more demanding question is whether a relation calibrated from inactive galaxies has the appropriate slope, normalization, and intrinsic scatter to provide reliable \mbh\ estimates for individual AGN. The presence of a positive correlation does not by itself establish that the relation has sufficient predictive accuracy for this purpose. Systematic errors in \mbh\ propagate directly into Eddington ratios and could bias comparisons among AGN classes and population-level inferences about black hole growth.

A substantial literature has investigated the \msigma\ relation in active galaxies, particularly in local type 1 AGN with \mbh\ estimates
obtained from RM or single-epoch spectroscopy \citep[e.g.,][]{Greene2006,Shen2008,Woo2010,Bennert2011,Woo2013,Bennert2021,Caglar2023,Winkel2025,Bennert2026}. Many of these studies find flatter slopes and larger scatter than relations derived from inactive galaxies \citep[e.g.,][]{Gueltekin2009,Kormendy2013}. Some authors argue that the active- and inactive-galaxy relations can be reconciled after accounting for their different \mbh\ dynamic ranges, host-galaxy properties, sample-selection effects, and regression biases. These considerations are important because local AGN samples generally contain lower-mass black holes and span a narrower range in \mbh\ than the dynamically measured inactive-galaxy samples \citep[e.g.,][]{Shankar2016}.

The existing work does not, however, establish that an inactive-galaxy relation provides reliable individual black hole masses across the broader AGN population. Most active-galaxy studies are dominated by optically selected type 1 AGN, and many use RM or single-epoch masses whose normalization is itself connected to an adopted host-galaxy relation. They also differ in their black hole mass range, accretion rate distribution, host morphology, and treatment of selection effects. Relatively few studies contain substantial samples of obscured AGN and apply the same mass estimator homogeneously across optical AGN classes. Agreement between active and inactive galaxies in an ensemble average therefore does not necessarily demonstrate that the standard \msigma\ relation is an accurate predictor of \mbh\ for an individual obscured AGN.

There are observational indications that applying inactive-galaxy relations to type 2 AGN can produce systematically biased masses. \citet{Greene2016} found that standard \msigma\ relations tend to overpredict the dynamically measured black hole masses of megamaser galaxies by 0.6 dex on average. \citet{Ricci2017b} reached a similar conclusion (with a discrepancy of 0.9 dex) using type 2 AGN whose BLRs are detected in the infrared, allowing single-epoch virial mass estimates. \citet{Gliozzi2024} carried out a systematic comparison of several indirect mass estimators for a volume-limited, hard-X-ray-selected sample with available dynamical masses. They found that masses inferred from a standard inactive-galaxy \msigma\ relation are systematically higher than the dynamical values by about one order of magnitude, whereas the X-ray scaling method yields masses consistent with dynamical \mbh\ values. These results motivate a test using a sample that includes both unobscured and obscured AGN and a mass estimator whose application does not require a directly visible BLR or the prior adoption of a host-galaxy scaling relation.

One of the major challenges of statistical AGN studies is obtaining observations of a complete and representative sample, relatively free of bias.   We base this current work on the local 12\,$\mu$m AGN sample. The sample is selected through its mid-infrared emission and is therefore insensitive to optical obscuration \citep{Rush1993}. This is a uniform all-sky bolometric-flux-limited selection that is not dependent on the presence of directly visible broad optical lines. Down to the bolometric flux limit of $2\times10^{-10}\,\mathrm{erg\,s^{-1}\,cm^{-2}}$ \citep{Rush1993}, it therefore includes a relatively unbiased sample of Seyfert~1 galaxies, whose optical spectra show broad permitted emission
lines; Seyfert~2 galaxies, whose optical spectra lack such broad lines; and HBLR galaxies, which are optically classified as Seyfert~2 systems but reveal broad emission lines in polarized light. A key advantage of the 12\,$\mu$m AGN sample is that it includes all the brightest AGN in the sky, and therefore has the most complete set of good multi-wavelength supporting data, with much higher detection rates than other AGN samples. This brightness is particularly important since the present study requires suitable \nustar\ observations, sufficient X-ray signal-to-noise ratio (S/N), host-galaxy measurements, and an accretion state within the range over which the X-ray scaling method is calibrated. 

We take advantage of the good-quality \nustar\ observations available for a large fraction of the 12\,$\mu$m sample to perform a systematic and homogeneous analysis of their hard-X-ray spectra. We use the X-ray scaling method to estimate \mbh\ for AGN accreting at moderate to high rates, including both unobscured and heavily obscured systems. Because the method does not require a visible BLR, it places Seyfert~1, Seyfert~2, and HBLR galaxies on a common mass scale. This enables us to measure their Eddington ratios and to investigate whether the presence of directly or indirectly detected broad lines is associated with intrinsic differences in accretion rate.

Our primary objective is to determine whether the \msigma\ relation of this mid-infrared-selected AGN sample is consistent with relations calibrated from inactive galaxies and whether it has sufficient predictive precision to serve as an indirect black hole mass estimator. We also examine the \mbh--$M_\star$ relation, where $M_\star$ is the stellar mass of the entire galaxy, and compare it with recent active- and inactive-galaxy determinations. Finally, we compare the distributions of \mbh\ and Eddington ratio \laedd=\lbol/\ledd\ (where \lbol\ is the bolometric luminosity and $L_\mathrm{Edd}=1.3\times10^{38}\, (M_\mathrm{BH}/M_\odot)\,\mathrm{erg\,s^{-1}}$ the Eddington luminosity) of Seyfert~1, Seyfert~2, and HBLR galaxies. Extending reliable black hole mass and accretion rate measurements to type 2 AGN is particularly important because they represent a major, and over much of the AGN luminosity function dominant, component of the accreting black hole population.

The structure of the paper is as follows. In Section~2, we describe the sample selection and the reduction of the \nustar\ data. In Section~3, we present the X-ray spectral analysis, determine the \mbh\ values with the X-ray scaling method, and compare those values with dynamical measurements. In Section~4, we compare \mbh\ and the Eddington ratio  of Seyfert~1, Seyfert~2, and HBLR galaxies and investigate the \msigma\ and  \mbh--$M_\star$ relations. We discuss the implications of the results in Section~5 and summarize our conclusions in Section~6. In the Appendix, which contains our tables, we also summarize the main characteristics, typical uncertainties, and limitations of the X-ray scaling method.

Throughout the paper, we adopt the cosmological parameters $H_0 = 70\,\mathrm{km~s^{-1}~Mpc^{-1}}$, $\Omega_\mathrm{m} = 0.28$, and $\Omega_{\Lambda} = 0.72$, consistent with the Wilkinson Microwave Anisotropy Probe (WMAP) nine-year measurements. The figures in this paper were created with the Tool for OPerations on Catalogues And Tables (TOPCAT) \footnote{\url{https://www.star.bristol.ac.uk/mbt/topcat/}}.

%%%%%%%%%%%%%%%%%%%%%%%%%%%%%%%%%%%%%%%%%%
%%%% SECTION 2: OBSERVATIONS 
%%%%%%%%%%%%%%%%%%%%%%%%%%%%%%%%%%%%%%%%%%
\section{Observations}
\label{sec2:Observations}
%%%%%%%%%%%%%%%%%%%%%%%%%%%%%%%%%%%%%%%%%%
%%%% SEC 2.1 SAMPLE SELECTION 
%%%%%%%%%%%%%%%%%%%%%%%%%%%%%%%%%%%%%%%%%%
\subsection{Sample selection}
\label{sec2.1:sample}
The original $12~\mu m$ AGN sample, selected from the IRAS catalog by \citealt{Spinoglio1989} and extended by \citealt{Rush1993}, contained 117 AGN divided into Seyfert 1 galaxies, Seyfert 2 galaxies, HBLR galaxies, and low-ionization nuclear emission-line galaxies (LINERs).

The chosen sample is essentially a bolometric flux-limited survey outside the galactic plane, and therefore largely unbiased, given the empirical fact that galaxies emit an approximately constant fraction of their total bolometric luminosity at $12\, \rm{\mu m}$ \citep{Spinoglio2024}. This fraction is $\sim 15\%$ for AGN, including Seyfert type 1s, type 2s, and quasars  \citep{Spinoglio1989}, and $7\%$ for normal and starburst galaxies, independent of star formation activity \citep{Spinoglio1995}.

This sample was selected from the 12\,$\mu$m flux-limited survey of 893 galaxies \citep{Rush1993} extracted from the IRAS Faint Source Catalog, Version 2 \citep[FSC-2;][]{Moshir1990} with the classification of galaxy nuclear activity with catalogs of active galaxies available at the time of the selection \citep{Veron1991, Hewitt1991, Helou1991}. 
Its 117 AGN consist of 48 Seyfert type 1s, 27 HBLRs, 30 Seyfert type 2s, and 10 LINERs, and two other galaxies, one classified as ``non-Seyfert'' (NGC\,1056) and one as a starburst galaxy (NGC\,6810). 

An important reason why this sample was chosen, besides the fact that it is representative of the AGN population of the local universe, is the availability of accurate {\it nuclear} bolometric luminosities for each object in the sample. These refer to the AGN, that is, they are corrected for the starlight contamination of the host galaxy \citep{Spinoglio2024}.

The bolometric luminosities of the 12\,$\mu$m AGN sample were computed through the integration of 10 photometric bands, from low to high frequencies: radio 8.4\,GHz, mid-IR 12\,$\mu$m, 7\,$\mu$m, 5.8\,$\mu$m, and 4.5\,$\mu$m bands, near-IR 2.2\,$\mu$m, the near ultraviolet (NUV) Galaxy Evolution Explorer \citep[GALEX,][]{Morrissey2007} band at an effective wavelength of 2316\,\AA, the far ultraviolet (FUV) GALEX band at 1539\,\AA, and the 2--10\,keV and 14--195\,keV hard X-ray bands \citep{Spinoglio2024}.
 From the available observations, the nuclear fluxes of the AGN were extracted, excluding the galactic component, which at some frequencies can be dominant. Sub-arcsecond observations at 12\,$\mu$m were used to correct the mid-IR band observations; the K-band nuclear flux was derived from Keck and 2MASS observations as reported and fully explained in \citet{Spinoglio2022}, and the ultraviolet fluxes were derived using the GALEX point spread functions. On the other hand, the 8.4\,GHz flux densities are already detecting the nuclei at sub-arcsecond resolution, and the hard X-ray emission is intrinsically produced mainly by the AGN.
 
As fully described in \citet{Spinoglio2024}, we used trapezoidal integration to calculate the total, or bolometric, luminosity for our AGN sample. 
The integration was performed in two separate spectral regions: the first interval is from the radio (8.4\,GHz band) to the FUV band, while the second interval includes only the two X-ray bands (2--10\,keV and 14--195\,keV). For those AGN where only one X-ray band flux was available, we computed the missing flux using the median spectral index between 2--10\,keV and 14--195\,keV for each AGN class.  The resulting bolometric luminosity is simply the sum of these two integrations.  No interpolation was performed between the UV and X-ray data because this spectral region is not observable and a power-law interpolation would likely overestimate the bolometric luminosity. Moreover, no extrapolation was made.

The full description of how the measurements were made to minimize the galaxy contribution as well as the uncertainties and limitations of the method used to compute the bolometric luminosities and build up the AGN spectral energy distributions can all be found in \citet{Spinoglio2024}.

In this work, we focus exclusively on AGN accreting at moderate--high levels (in other words, we excluded the LINERs) for two reasons: 1) to compare and contrast BH--galaxy relationships of genuine AGN with those obtained with inactive galaxies with low-accreting AGN reported in the literature, and 2) because the X-ray scaling method that we use to constrain \mbh\ is calibrated for AGN accreting at moderate--high level and cannot be safely applied to low-accreting sources. 

Of the original 117 galaxies, 96 have \nustar\ archival observations (specifically, 39 Seyfert 1s, 32 Seyfert 2s, and 25 HBLRs). After excluding sources without clear detections or with insufficient S/N for a meaningful X-ray spectral analysis (see details in the next section) or jet-dominated (3C 273), our final sample comprises 73 AGN (33 Seyfert 1s, 18 Seyfert 2s, and 22 HBLRs). The main properties of this sample are reported in Table \ref{tab:general_properties}.

%%%%%%%%%%%%%%%%%%%%%%%%%%%%%%%%%%%%%%%%%%
%%SEC 2.2 DATA REDUCTION SEC.
%%%%%%%%%%%%%%%%%%%%%%%%%%%%%%%%%%%%%%%%%%
\subsection{NuSTAR data reduction}
\label{sec2.2:data}
We processed all the observations with the \nustar\ data analysis script \textsc{nupipeline}, which calibrates and cleans the data with the calibration database CALDB\footnote{\url{https://heasarc.gsfc.nasa.gov/docs/heasarc/caldb/nustar/}} version 20200429. Utilizing the \textsc{nuproducts} script, for each AGN we extracted spectra and light curves for both focal plane modules FPMA and FPMB using a circular aperture centered on the source as extraction region and one in a nearby region of the same image to define the background; for both regions the radii range from 40 arcsec for dimmer sources to 60 arcsec for the brighter ones.

We report all the \nustar\ observations in Table \ref{tab:x-ray_properties} in the Appendix, where we list the effective exposure times (the actual time duration to collect X-ray photons, calculated after subtracting periods of instrumental dead time, Earth occultations, and high-background solar or particle flares), along with the 2--10 keV luminosities and the main spectral parameters derived from the X-ray spectral analysis, which is described in the next section.

%%%%%%%%%%%%%%%%%%%%%%%%%%%%%%%%%%%%%%%%%%
%%SEC 3 X-RAY SPECTRAL ANALYSIS & MBH ESTIMATE
%%%%%%%%%%%%%%%%%%%%%%%%%%%%%%%%%%%%%%%%%%
\section{X-ray spectral analysis and \mbh\ estimates}
\label{sec3:X-ray spectral}

%%%%%%%%%%%%%%%%%%%%%%%%%%%%%%%%%%%%%%%%%%
%%SEC 3.1 NUSTAR SPECTRAL ANALYSIS
%%%%%%%%%%%%%%%%%%%%%%%%%%%%%%%%%%%%%%%%%%
\subsection{NuSTAR spectral analysis}
\label{sec3.1:spectra}
The X-ray emission in actively accreting BH systems is thought to be produced in a hot corona by the Comptonization of seed photons produced in the accretion disk. To properly characterize the X-ray emission, especially in obscured sources, which represent the majority of our sample, it is important to accurately constrain the contributions of absorption and reflection.

To this end, we exploited the high-quality broadband X-ray spectra provided by \nustar\ and employed physically motivated models of the putative obscuring torus. As in our previous works \citep{Gliozzi2021,Gliozzi2024}, we used as baseline model 
\begin{verbatim}
phabs * (atable(Borus) + MYTZ*BMC)
\end{verbatim}
where the \texttt{phabs} model describes the Galaxy absorption. The \borus\ %table 
model accounts for the continuum scattering and fluorescent emission line components associated with the torus \citep{Balokovic2018}. The zeroth-order component of MYTorus, \texttt{MYTZ}, which is applied to the transmitted primary emission, parameterizes the absorption and Compton scattering using the Klein-Nishina cross section \citep{Murphy2009}.

The free parameters of the bulk motion Comptonization model \texttt{BMC} \citep{Titarchuk1997} are the normalization \nbmc\ (in units of luminosity divided by the distance squared), the spectral index $\alpha$ (where $\Gamma=\alpha+1$), the temperature of the seed photons $kT$, and $\log A$, which is related to the fraction of scattered seed photons $f$ by the relationship $A=f/(1 - f)$. 

To ensure the self-consistency of the torus model, we linked the spectral index  $\alpha$ of the Comptonization model to the $\Gamma$ value of the torus model and set \nbmc\ equal to the \borus\ power-law normalization divided by a factor of 30, based on the empirical relationship obtained in \citet{Gliozzi2021}.  

For AGN with relatively low absorption (in our sample this means most of the Seyfert 1s, which have $\log (N_{\mathrm{H}}/{\mathrm{cm^{-2}}})$ less than 23.5), we parameterized  the attenuation of the transmitted component with the \texttt{zphabs} model instead of \texttt{MYTZ}. A few sources required some additional components to the baseline model, for instance, an extra Gaussian line around 6.4 keV, or a (\texttt{const*bmc}) model that describes the fraction of primary emission that is directly scattered towards the observer by an optically thin ionized medium. 

We performed the X-ray spectral analysis using the \xspec\ \texttt{v.12.9.0} software package \citep{Arnaud1996}, and  the errors quoted on the spectral parameters represent the 1$\sigma$ confidence level. All spectra were binned with a minimum of 20 counts per bin  for the $\chi^2$ statistics to be valid.

%%%%%%%%%%%%%%%%%%%%%%%%%%%%%%%%%%%%%%%%%%
%%SEC 3.2 X-RAY SCALING MBH
%%%%%%%%%%%%%%%%%%%%%%%%%%%%%%%%%%%%%%%%%%
\subsection{X-ray-based \mbh\ estimates}
\label{sec3.2:mbh}
The X-ray scaling method estimates the mass of SMBHs in actively accreting AGN by scaling up the dynamical mass of stellar-mass BHs residing in well-studied X-ray binaries, which are used as reference sources. This method, developed for stellar-mass BHs by \citet{Shaposhnikov2009} was tested and successfully applied to type 1 AGN of the RM sample \citep{Gliozzi2011}, as well as to type 2 AGN with \mbh\ determined via water maser measurements \citep{Gliozzi2021}.  More recently, using a volume-limited sample of AGN detected at very hard X-ray energies (the BAT sample), \citet{Gliozzi2024} showed that the \mbh\ values determined with this method are more robust and reliable than those obtained with other indirect methods commonly used in the literature, such as the SE method \citep[e.g.,][]{Vestergaard2002}, the \msigma\ correlation \citep[e.g.,][]{Kormendy2013}, the fundamental plane for black hole activity \citep[e.g.,][]{Merloni2003}, and the X-ray variability method \citep[e.g.,][]{Akylas2022}.

The basic assumptions, uncertainties, and limitations of the X-ray scaling method are summarized in the Appendix. The \mbh\ values obtained with the X-ray scaling method along with their uncertainties and the Eddington ratio \laedd\ are listed in Table \ref{tab:BH_masses} in the Appendix.

To verify that the \mbh\ estimates obtained with the X-ray scaling method are reliable, for all the sources of the 12\,$\mu$m \nustar\ AGN sample we searched the literature and collected all available \mbh\ values obtained with dynamical methods: six megamaser disk measurements, one gas dynamics measurement, seven measurements from reverberation mapping based on direct modeling of the BLR,  and 14 standard RM measurements. The main difference between the two RM methods is the following. In the standard RM a sample-averaged virial factor $\langle f \rangle$ is introduced to account for the unknown geometry and kinematics of the BLR. On the other hand, in the direct modeling method (applicable only to datasets with higher S/N), the BLR geometry is determined by forward modeling velocity-resolved BLR data using a Bayesian code for AGN reverberation and modeling of emission lines (CARAMEL) introduced by \citet{Pancoast2011}. 

In order to increase the size of this dynamical subsample and improve the statistics for the comparison with the X-ray based \mbh\ values, we also included \mbh\ obtained with the SE method, which strictly speaking is a secondary indirect method. Although some recent studies suggest that SE values can be underestimated in sources with substantial absorption (see, e.g., \citealt{Mejia2022,Caglar2023,Gliozzi2025}) and overestimated in sources accreting at a high level (see, e.g., \citealt{Du2014,Martinez2019,Gliozzi2025,Woo2026}), the AGN with SE measurements in the 12\,$\mu$m sample are not heavily obscured or highly accreting and hence their \mbh\ values are likely reliable and consistent with the RM estimates. Including these 13 SE measurements, we have a total of 41 sources with dynamically constrained \mbh\ that can be compared to the X-ray based estimates.

All the \mbh\ values derived with dynamical methods and their respective uncertainties and references are listed in Table \ref{tab:BH_masses}. The only changes with respect to the values reported in the references are the estimates obtained with methods that directly depend on the source distances (namely, maser and gas dynamics), which were rescaled to be consistent with the distances used in this work. For RM and SE estimates we adopted systematic errors of 0.4 dex and 0.6 dex respectively, which better reflect the actual uncertainties of these methods (e.g., \citealt{Peterson2014,Mejia2022,Shen2024,Vestergaard2006}).

\begin{figure}[ht!]
   \centering
   \includegraphics[width=\hsize]{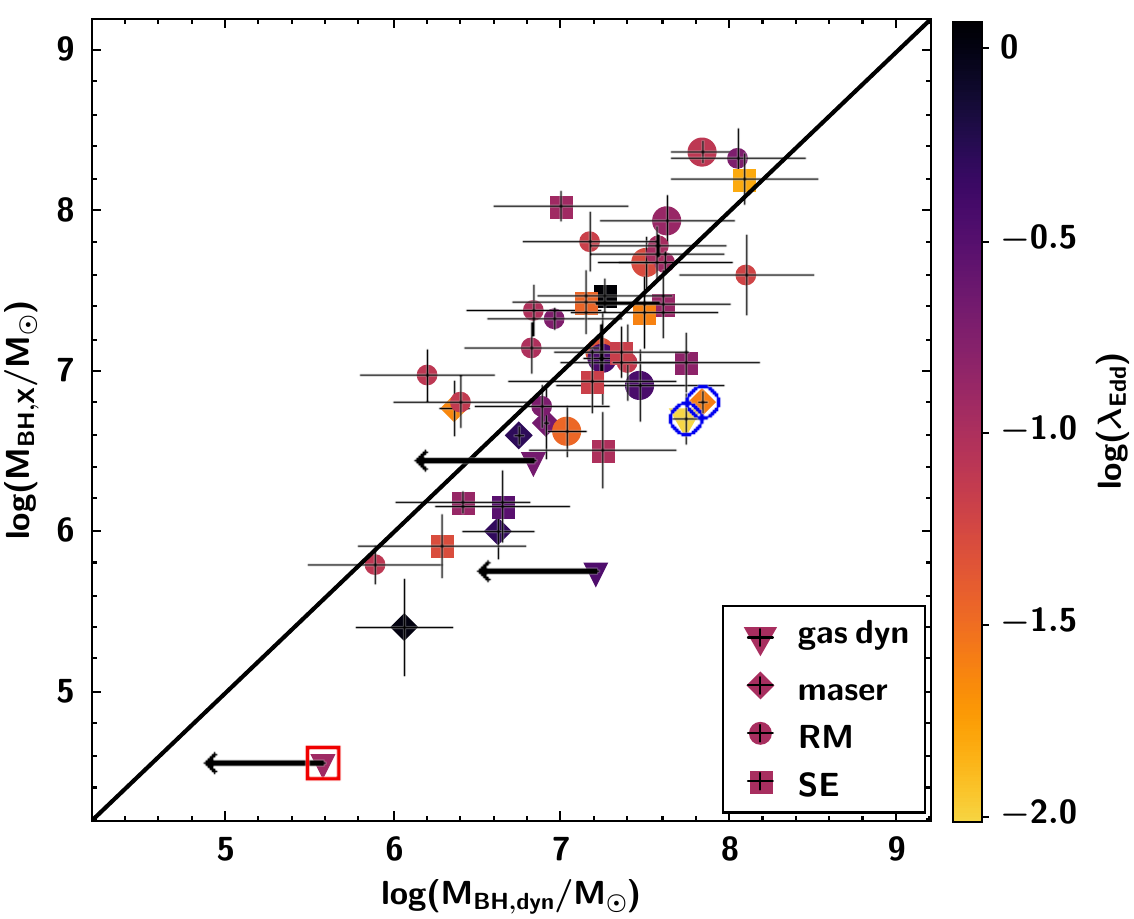}%
      \caption{log \mbh\ from  X-ray scaling  plotted vs. log \mbh\ from dynamical methods with a color coding scheme illustrating the level of the Eddington ratio $\lambda_\mathrm{Edd}$. The continuous black line indicates the one-to-one correlation. Downward-pointing triangles represent values obtained with gas dynamics, diamonds are maser values, squares are values obtained with the single epoch method. Larger circles indicate sources with direct modeling of the BLR, whereas the smaller circles show \mbh\ obtained with the standard RM. The error bars represent the uncertainties associated with the different methods and reported in Table \ref{tab:BH_masses}. The two blue open circles indicate NGC 1194 and NGC 7582 (the only sources not consistent within their uncertainties), whereas the open red square indicates NGC 5194, the outlier with the lowest \mbh.}
         \label{figMdMx}
   \end{figure}

Fig.~\ref{figMdMx} shows the \mbh\ values obtained with the X-ray scaling method ($M_\mathrm{BH,X}$) plotted versus the corresponding values derived with various dynamical methods  ($M_\mathrm{BH,dyn}$) with a color-coded scheme based on \laedd. 

A visual inspection of this figure indicates the overall good agreement between values derived with the X-ray scaling method and the dynamical ones, as confirmed by a Spearman rank coefficient of $\rho=0.66$ ($P_\mathrm{S}=2.3\times10^{-6}$) and an orthogonal distance regression (ODR; \citealt{Spinoglio2022}) fit: $\log(M_\mathrm{BH,dyn}/M_\odot)=(0.85\pm0.72)+(0.90\pm0.10)\times\log(M_\mathrm{BH,X}/M_\odot)$. NGC 1194 and NGC 7582, which are not formally consistent with the dynamical values, are included in this fit. In the remainder of the paper, for these two sources we do not use the \mbh\ values obtained with the X-ray scaling method but instead their dynamically derived mass estimates.

The strong positive correlation consistent with the one-to-one correspondence was confirmed by a linear regression carried out with the \textsc{linmix\textunderscore err} routine \citep{Kelly2007}, which uses a Bayesian approach and accounts for errors both on the x-axis and on the y-axis.  This yields $\log(M_\mathrm{BH,dyn}/M_\odot)=(0.22\pm1.36)+(0.96\pm0.19)\times\log(M_\mathrm{BH,X}/M_\odot)$ and an intrinsic scatter of $\epsilon=0.23\pm0.08$ dex. 

To further quantitatively test the consistency between $\log(M_\mathrm{BH,dyn}/M_\odot)$ and $\log(M_\mathrm{BH,X}/M_\odot)$, for each source we computed their difference using the quantity $\Delta\log(M_\mathrm{BH}/M_\odot)/\sigma_\mathrm{diff}$, where $\Delta\log(M_\mathrm{BH}/M_\odot)=\lvert \log(M_\mathrm{BH,dyn}/M_\odot)-\log(M_\mathrm{BH,X}/M_\odot)\rvert$ and $\sigma_\mathrm{diff}=\sqrt{\sigma_\mathrm{dyn}^2+\sigma_\mathrm{X}^2}$, where $\sigma_\mathrm{dyn}$ and  $\sigma_\mathrm{X}$ are the uncertainties on $\log(M_\mathrm{BH,dyn}/M_\odot)$ and $\log(M_\mathrm{BH,X}/M_\odot)$, respectively. In other words, the quantity $\Delta\log(M_\mathrm{BH}/M_\odot)/\sigma_\mathrm{diff}$ measures the difference between the logarithms of the dynamical black hole mass and the X-ray \mbh\ in terms of sigmas.
For the whole sample of 41 sources with dynamical mass estimates, the average value of $\Delta\log(M_\mathrm{BH}/M_\odot)/\sigma_\mathrm{diff}$ is 1.5 and the median 0.8, confirming the good agreement between the two \mbh\ estimates. The only two exceptions are NGC 1194 and NGC 7582 (indicated by the blue circles in Fig.~\ref{figMdMx}), whose $\log(M_\mathrm{BH,dyn})$ values are significantly larger than $\log(M_\mathrm{BH,X})$.
%_____________________________________________________________________________________
%                        figure 1 log(MBH_dyn) vs log(MBH_X) as large as the column width
%-----------------------------------------------------------------------------------------
%%_____________________________________________________________

Based on these tests, we confirm that the X-ray scaling method provides \mbh\ values consistent with those obtained from dynamical methods regardless of the level of obscuration. Therefore, for the rest of the paper we use \mbh\ estimates obtained with the X-ray scaling method, which depend on the spectral parameters derived in our homogeneous and systematic spectral analysis of the \nustar\ data. The only two exceptions are NGC 1194 and NGC 7582, for which we will use the dynamical values, because these sources are relatively low accreting (\laedd$\sim 10^{-2}$) and the X-ray scaling method may not be reliable in that accretion regime \citep{Jang2014}.

%%%%%%%%%%%%%%%%%%%%%%%%%%%%%%%%%%%%%%%%%%
%%%% SECTION 4: RESULTS
%%%%%%%%%%%%%%%%%%%%%%%%%%%%%%%%%%%%%%%%%%
\section{Results}
\label{sec4:Results}

%%%%%%%%%%%%%%%%%%%%%%%%%%%%%%%%%%%%%%%%%%
%%%%%% SEC 4.1 MBH-lambda_Edd Distributions 
%%%%%%%%%%%%%%%%%%%%%%%%%%%%%%%%%%%%%%%%%%
\subsection{\mbh--\laedd\ distributions}
\label{sec4.1:mbhlaedd}

Knowing \mbh\ (see Sect. \ref{sec3.2:mbh} and Table \ref{tab:BH_masses}) and the bolometric luminosity (Sect. \ref{sec2.1:sample} and Table \ref{tab:general_properties}) for all objects of our restricted $12~\mu m$ AGN sample, we compare the three AGN subclasses (Seyfert 1s, Seyfert 2s, and HBLRs) based on their fundamental properties, \mbh\ and \laedd. For the uncertainties on log(\laedd) we assume 0.2 dex, which are of the order of  the uncertainties of $\log(L_\mathrm{bol})$ obtained integrating AGN broadband SEDs \citep[e.g.,][]{Lusso2012} and the  uncertainties of the $\log(M_\mathrm{BH}/M_\odot)$ values obtained with the X-ray scaling method (the average and median for this sample are 0.19 dex and 0.18 dex, respectively).

\begin{figure}[ht!]
   \centering
   \includegraphics[width=\hsize]{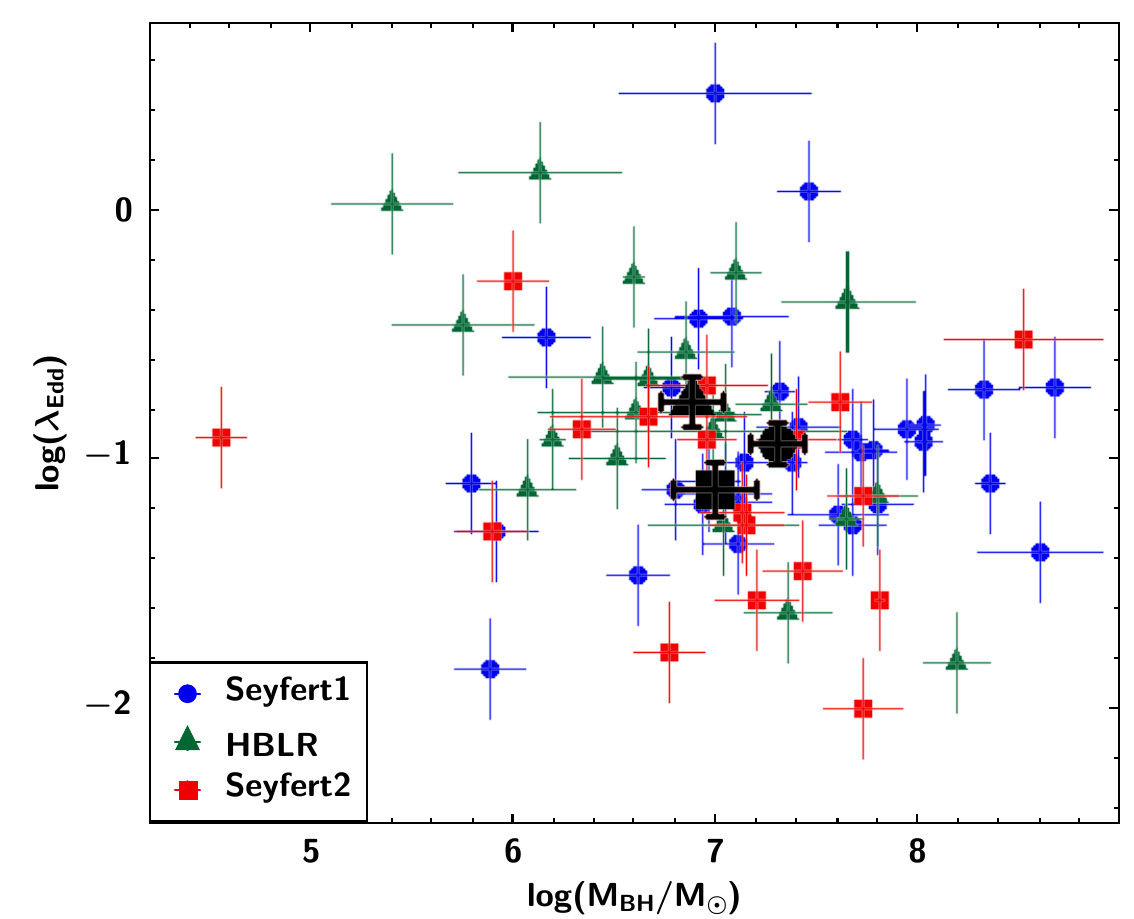}%{loglaEdM_Feb26B1.pdf}
      \caption{log(\laedd) plotted versus log(\mbh) for the Seyfert 1 (blue circles), Seyfert 2 (red squares), and HBLR galaxies (green triangles). For each colored symbol, the horizontal error bars represent the uncertainties on log(\mbh) and the vertical error bars of 0.2 dex are the uncertainties on  log(\laedd) used in our analysis. The corresponding larger black symbols represent the averages for the three AGN subclasses along with their respective uncertainties $\sigma/\sqrt{n}$, where $n$ is the number of objects in each AGN subclass.  }
         \label{fig_Mlaedd}
   \end{figure}

This comparison is illustrated in Fig.~\ref{fig_Mlaedd}, which shows a substantial overlap among the \mbh\ distributions of the three AGN subclasses, with HBLRs having the lowest average value ($\log(M_\mathrm{BH})=6.88\pm0.15$) compared to Seyfert 1s and Seyfert 2s ($\log(M_\mathrm{BH})=7.31\pm0.13$ and $\log(M_\mathrm{BH})=7.00\pm0.21$, respectively). Assessing the quantitative difference with a Student's t-test reveals that the \mbh\ distribution for HBLRs differs from the corresponding Seyfert 1 distribution at the 2.2$\sigma$ level ($P_t=3\times10^{-2}$), whereas there is no significant difference between the HBLR and the Seyfert 2 mass distributions.
%_____________________________________________________________________________________
%                        figure 2 log(lambda_Edd) vs log(MBH) as large as the column width
%-----------------------------------------------------------------------------------------
%
%%_____________________________________________________________

The same figure suggests a possible difference between the log(\laedd) averages of HBLRs and Seyfert 2s ($-0.77\pm0.10$ and $-1.11\pm0.10$, respectively). This is confirmed by a Student's t-test at the 2.3$\sigma$ level ($P_t=2.7\times10^{-2}$), whereas there is no significant difference between the HBLR and the Seyfert 1 accretion rate distributions.

For completeness, we also performed a Kolmogorov--Smirnov (K--S) test comparing the \mbh\ and \laedd\ distributions of the three AGN subclasses. The only suggestive evidence for a difference is that between the \laedd\ distributions of the HBLRs and the Seyfert 2s (D = 0.36, $P_{KS}=1.4\times10^{-1}$).

In summary, our analysis shows marginal evidence at the 2.3$\sigma$ level that the two AGN classes of Seyfert 2 and  HBLR galaxies, which are characterized by similar high levels of  obscuration, appear to be intrinsically different, with the latter class having substantially higher \laedd\ values, which are instead consistent with Seyfert 1 accretion levels. 
The intrinsic similarity between obscured HBLR galaxies and unobscured Seyfert 1 galaxies is consistent with the AGN unification paradigm \citep{Antonucci1993}. These findings also confirm those of \citet{Tommasin2010}, based on the mid-IR spectroscopy of the 12\,$\mu$m AGN sample, who found that Seyfert galaxies are divided into the two main classes of type 1 AGN, which includes Seyfert 1s and HBLRs, and type 2 AGN, which contains only Seyfert 2s with no detectable broad lines in either direct or polarized spectroscopy.

%%%%%%%%%%%%%%%%%%%%%%%%%%%%%%%%%%%%%%%%%%
%%%%%% SEC 4.2 MBH-SIGMA CORRELATION 
%%%%%%%%%%%%%%%%%%%%%%%%%%%%%%%%%%%%%%%%%%
\subsection{\msigma correlation}
\label{sec4.2:Msigma}
\begin{figure}[ht!]
   \centering
   \includegraphics[width=\hsize]{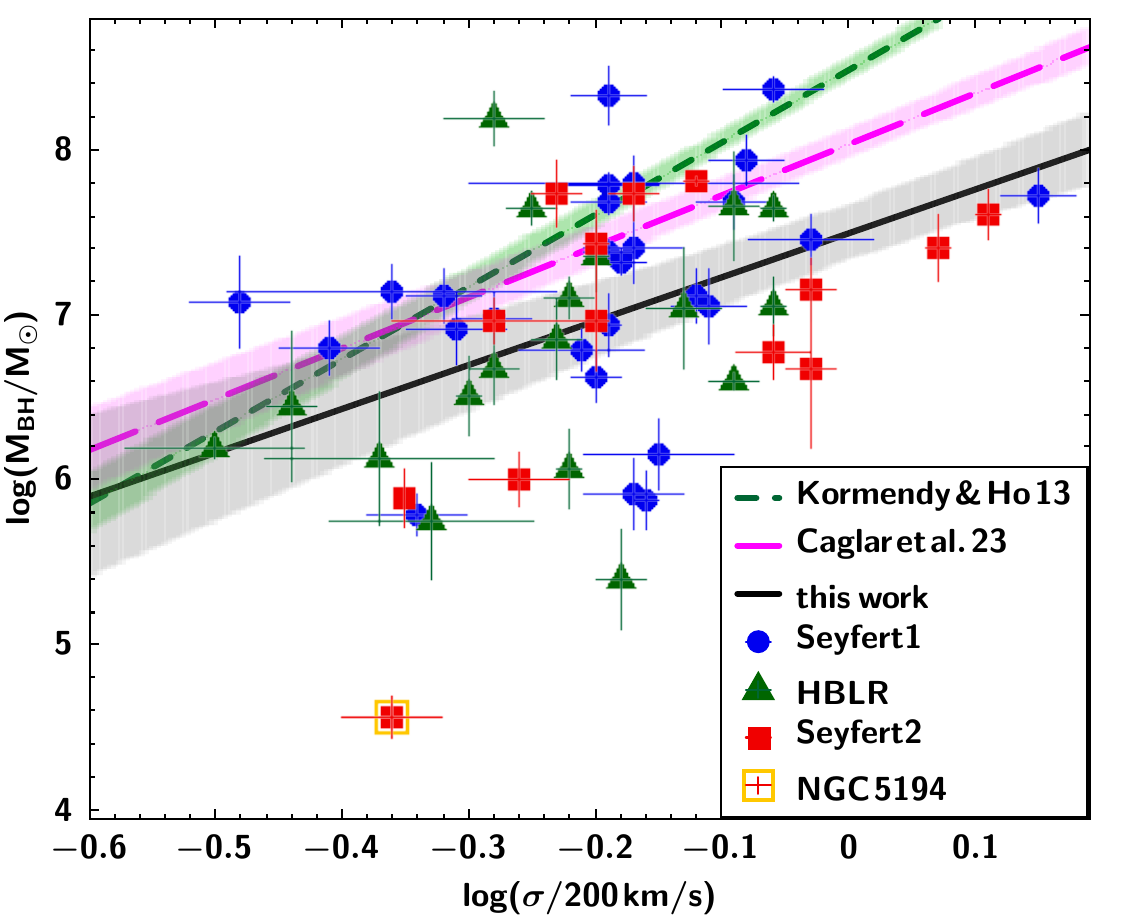}%{Msig60_May26.pdf}
      \caption{$\log$(\mbh) plotted versus $\log(\sigma_\star)$ for the Seyfert 1 (blue circles), Seyfert 2 (red squares), and HBLR galaxies (green triangles). The uncertainties on the bulge velocity dispersion are reported in Table \ref{tab:general_properties} along with the respective references. The black continuous line represents the linear best fit, the green short-dashed line indicates the best fit found by \citet{Kormendy2013} using inactive galaxies with classical bulges, and the magenta long-dashed line shows the best fit  obtained by \citet{Caglar2023} using a large sample of type 1 AGN. The shaded regions account for the errors on the intercepts and slopes of the best-fit linear models.}
         \label{fig-msigma}
   \end{figure}

Fig.~\ref{fig-msigma} illustrates the \msigma\ correlation for the \nustar\ 12\,$\mu$m sample including Seyfert 1, Seyfert 2, and HBLR galaxies. There is an apparent positive trend, confirmed by a Spearman rank linear correlation coefficient of $\rho=0.45$ ($P_S=4.1\times10^{-4}$), without any clear difference in the distributions of the three types of galaxies in this plane. A visual inspection of the same figure also reveals a substantial scatter (rms = 0.63 or 0.68 dex, depending on the inclusion of the outlier NGC 5194, which has the lowest \mbh\ of the sample by nearly an order of magnitude). 

%_____________________________________________________________________________________
%                        figure 3 log(MBH) vs log(sigma/200km/s) as large as the column width
%-----------------------------------------------------------------------------------------

The best linear fit, obtained with the \textsc{linmix\textunderscore err} routine \citep{Kelly2007} is
$$\log(M_\mathrm{BH}/M_\odot)=(7.50\pm0.17)+(2.66\pm0.77)\times \log(\sigma_\star/200\,\mathrm{km~s^{-1}}),$$ with an intrinsic scatter of $\epsilon=0.45\pm0.10$.
If we exclude NGC 5194, the best linear fit is $\log(M_\mathrm{BH}/M_\odot)=(7.46\pm0.16)+(2.26\pm0.70)\times \log(\sigma_\star/200\,\mathrm{km~s^{-1}})$, with an intrinsic scatter of $\epsilon=0.37\pm0.08$.

In order to strengthen this finding, we show that the sample used to derive the \msigma\ correlation is representative of the whole 12\,$\mu$m AGN parent sample. Our sample is composed of 59 AGN (all the AGN that have measured $\sigma_{\star}$ and good-quality X-ray data to determine \mbh, i.e., 27 Seyfert 1, 14 Seyfert 2, and 18 HBLR galaxies). On the other hand, the entire 12\,$\mu$m AGN parent sample (excluding the LINERs and the two non-Seyfert galaxies) comprises 104 objects (namely, 48 Seyfert 1, 29 Seyfert 2, and 27 HBLR galaxies). 

First of all, the fraction of objects of the same class is similar: we use 56\% of the total of Seyfert 1, 48\% of the total of Seyfert 2, and 66\% of the total of HBLR galaxies.

\begin{figure}[ht!]
   \centering
   \includegraphics[width=\hsize]{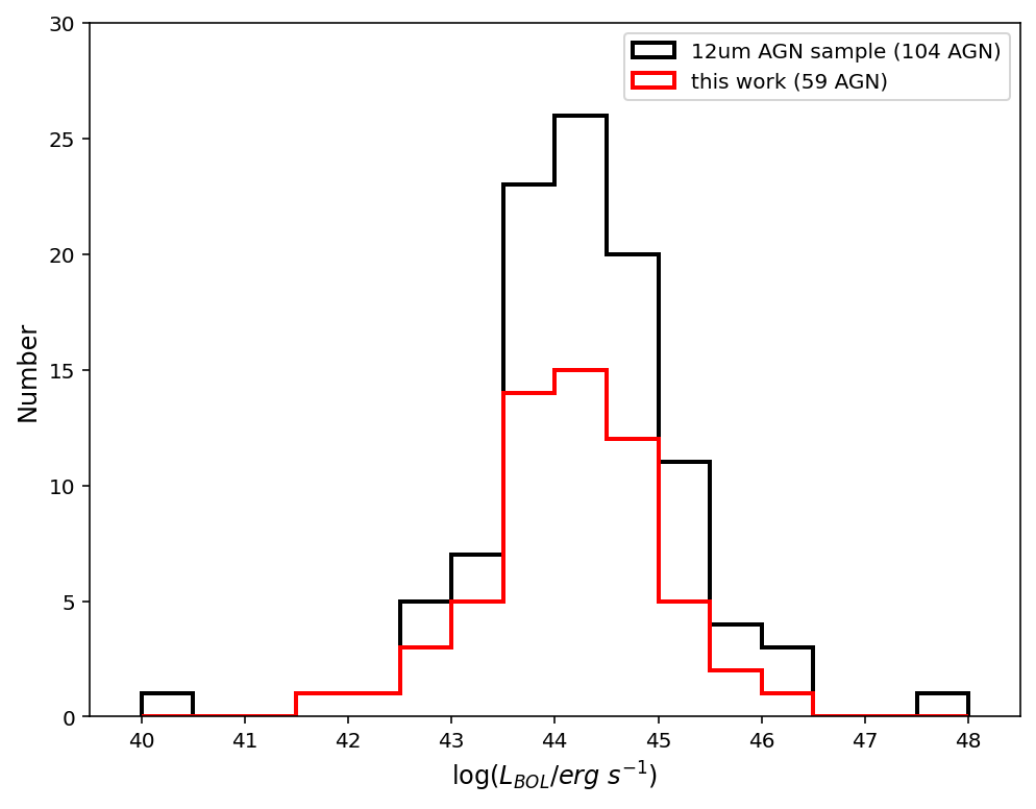}%{luminosity_histogram.pdf}
      \caption{The histogram of the bolometric luminosity $\log(L_\mathrm{bol}$) of the two samples. In black the complete 12\,$\mu$m AGN sample from which the LINERs have been subtracted, containing 104 AGN and in red the sample used in this work, with good \nustar\ spectral data, containing 59 AGN.
     }
         \label{fig-lumhisto}
   \end{figure}

To test whether there is a quantitative difference, we compared three fundamental properties, the bolometric luminosity, the stellar mass content of the galaxy, and the star formation rate of the 12\,$\mu$m AGN sample with good \nustar\ spectral data (59 AGN) with the parent sample of 104 AGN.
In Fig.~\ref{fig-lumhisto} we show the histogram of the bolometric luminosity $\log(L_\mathrm{bol}$) of the two samples. The statistics is as follows: the parent sample of 104 AGN is characterized by a median logarithmic bolometric luminosity and standard deviation of $\log(L_\mathrm{bol}/{\rm erg~s^{-1}})$ = 44.28 $\pm$ 0.97, while the used sample of 59 AGN has:  $\log(L_\mathrm{bol}/{\rm erg~s^{-1}})$ = 44.16 $\pm$ 0.85. Moreover, a Kolmogorov--Smirnov test gives a D-value of D = 0.0683 and a p-value of 0.9878, showing that there is no statistical difference between the two distributions.

%________________________________________________________________________________________
%                        figure 4 luminosity histogram
%-----------------------------------------------------------------------------------------
%-----------------------------------------------------------------------------------------

\begin{figure}[ht!]
   \centering
   \includegraphics[width=\hsize]{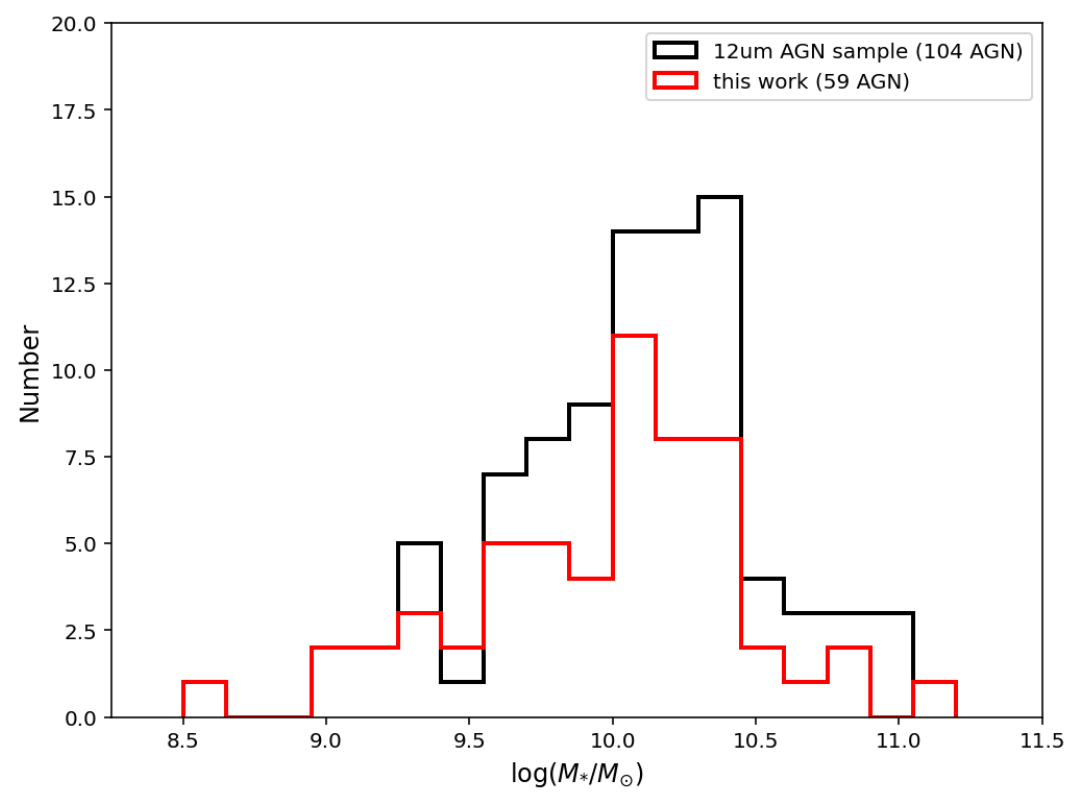}%{histogram_stellarmass.pdf}
      \caption{The histogram of the stellar mass $\log$(M$_{*}$/M$_{\odot}$) of the two samples. In black the complete 12\,$\mu$m AGN sample from which the LINERs have been subtracted, containing 104 AGN and in red the sample used in this work, with good \nustar\ spectral data, containing 59 AGN.
     }
         \label{fig-masshisto}
   \end{figure}

A similar result is found if we compare the two samples in terms of the stellar masses of the host galaxies of the AGN, as can be seen in the histogram in Fig.~\ref{fig-masshisto}. The parent sample of 104 AGN is characterized by a median logarithmic stellar mass and standard deviation of $\log$(M$_{*}$/M$_{\odot}$) = 10.13 $\pm$ 0.49, while the used sample of 59 AGN has  $\log$(M$_{*}$/M$_{\odot}$) = 10.07 $\pm$ 0.49. Moreover, a Kolmogorov--Smirnov test gives a D-value of D = 0.0934 and a p-value of 0.8822, showing that, also in this case, there is no statistical difference between the two distributions.

\begin{figure}[ht!]
   \centering
   \includegraphics[width=\hsize]{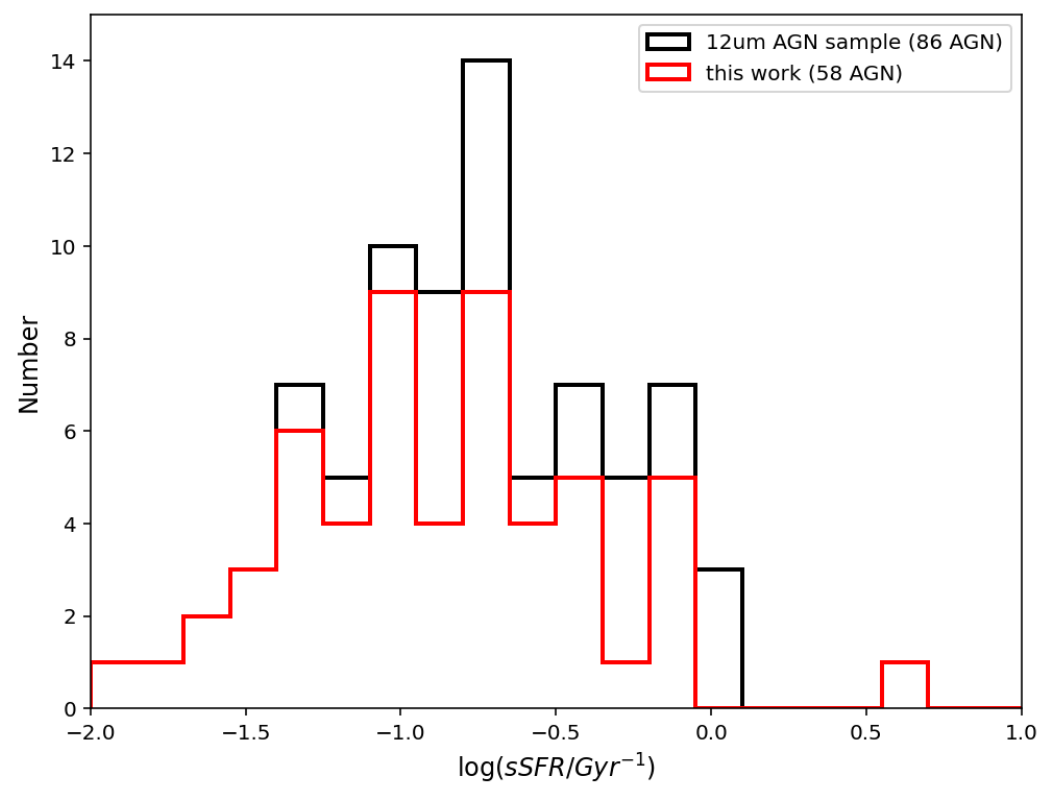}
      \caption{The histogram of the specific star formation rate $\log(\mathrm{sSFR}/\mathrm{Gyr^{-1}})$ of the two samples. In black the complete 12\,$\mu$m AGN sample from which the LINERs have been subtracted, containing 86 AGN and in red the sample used in this work, with good \nustar\ spectral data, containing 58 AGN.
     }
         \label{fig-sfrhisto}
   \end{figure}

Finally, similar conclusions are reached when we compare the two samples in terms of the specific star formation rate $sSFR=SFR/M_{*}$ of the host galaxies, as shown in the histogram in Fig.~\ref{fig-sfrhisto}. In this case, a Kolmogorov--Smirnov test gives a D-value of 0.0986 and a p-value of 0.8469, confirming that there is no statistical difference between the two distributions.

%________________________________________________________________________________________
%                        figure 5 M* histogram
%-----------------------------------------------------------------------------------------
%-----------------------------------------------------------------------------------------

The \msigma\ correlation, found for the 12\,$\mu$m sample (Fig.~\ref{fig-msigma}), is significantly different from that obtained by \citet{Kormendy2013}, using inactive nearby galaxies with classical bulges. Our correlation is characterized by a much larger dispersion and systematically lower M$_\mathrm{BH}$ for a given $\sigma_{\star}$ compared to \citet{Kormendy2013}. This difference is further discussed in Section~\ref{sec5.3:Mbh_sigma}.
%________________________________________________________________________________________
%                        figure 6 SFR histogram
%-----------------------------------------------------------------------------------------
%-----------------------------------------------------------------------------------------

%%%%%%%%%%%%%%%%%%%%%%%%%%%%%%%%%%%%%%%%%%
%%%%%% SEC 4.3 MBH-MGAL CORRELATION 
%%%%%%%%%%%%%%%%%%%%%%%%%%%%%%%%%%%%%%%%%%
\subsection{\mbh-M$_\star$ correlation}
\label{sec4.3:MMgal}

The complete details of the method used to compute the stellar masses of the host galaxies of the complete 12$\mu$m AGN sample are reported in \citet{Spinoglio2022}. Here we only give a summary of this method.
The stellar mass of the host galaxy of each AGN in the 12\,$\mu$m sample was derived from the K-band photometry: the galaxy stellar K-band flux was obtained by subtracting the nuclear flux from the total flux of each galaxy.  
Once transformed into the stellar galaxy luminosity, it was converted into stellar mass using the mass-to-light ratio in the K-band of 0.54 \citep{McGaugh2014}. The mean error in the K magnitudes is 0.032 mag, with a standard deviation $\sigma$ = 0.017 mag, which corresponds to a relative statistical error in the determination of the stellar mass of $\sim 3 \%$. However, given the uncertainty on the mass-to-light ratio, we conservatively assumed an uncertainty of 0.2 dex on $\log(M_\star/M_\odot)$ in our statistical analysis.

\begin{figure}[ht!]
   \centering
   \includegraphics[width=\hsize]{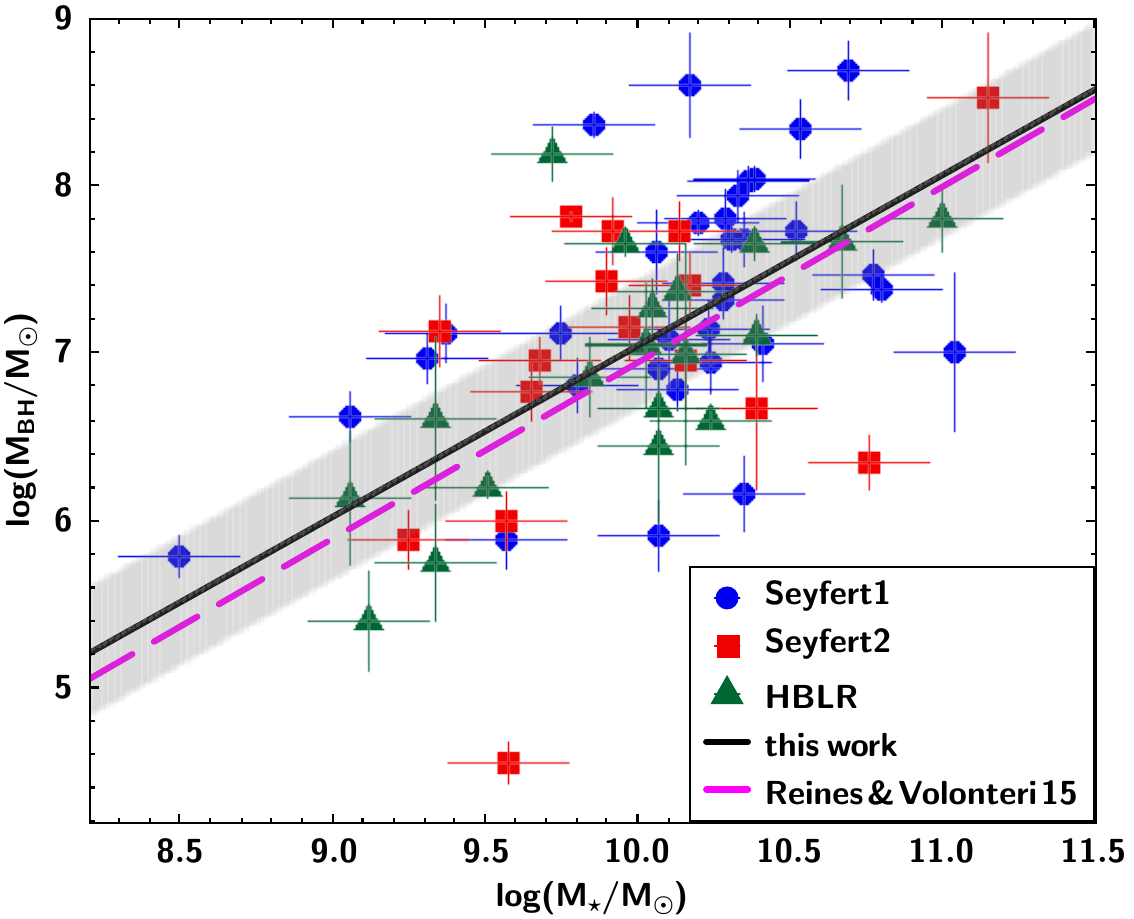}%
      \caption{Black hole mass in logarithmic form $\log$(\mbh) plotted versus the logarithm of the stellar mass $\log(M_{\star})$ for the Seyfert 1 (blue circles), Seyfert 2 (red squares), and HBLR galaxies (green triangles). The black continuous line represents the linear best fit and the magenta dashed line indicates the best fit found by \citet{Reines2015} using a large sample of type 1 AGN. The shaded region represents the 0.39 dex intrinsic scatter around our best-fit linear model. }
         \label{fig-MM*}
   \end{figure}

The correlation between \mbh\ and the galaxy stellar mass $M_\star$ is shown in Fig.~\ref{fig-MM*} and formally confirmed by a nonparametric Spearman analysis with a coefficient $\rho=0.51$ and associated null probability  $P_S=6.7\times10^{-6}$ (excluding NGC 5194, $\rho=0.52$ and $P_S=5.1\times10^{-6}$).
%_____________________________________________________________________________________
%                        figure 6 log(MBH) vs log(M_gal) as large as the column width
%-----------------------------------------------------------------------------------------
%-----------------------------------------------------------------------------------------
 
The $M_{BH}$--$M_{\star}$ correlation shown in Fig.\,~\ref{fig-MM*} has a best linear fit from \textsc{linmix\textunderscore err} of  
\noindent
 $$\log(M_\mathrm{BH}/M_\odot)=(-3.15\pm2.02)+(1.02\pm0.20)\times \log(M_\star/M_\odot) $$
\noindent
with intrinsic scatter $\epsilon=0.39\pm0.08$ dex. If NGC 5194 is excluded, $\log(M_\mathrm{BH}/M_\odot)=(-2.45\pm1.83)+(0.96\pm0.18)\times \log(M_\star/M_\odot)$, with $\epsilon=0.32\pm0.07$ dex. 
Once more, there appear to be no differences among the three subclasses of AGN in the 12 $\mu  $m sample, and the objects are distributed with a substantial scatter around the best-fit model. 

For comparison, in Fig.~\ref{fig-MM*}, we also show the best linear fit  found by \citet{Reines2015} using a large sample (341 objects) of relatively nearby optically selected type 1 AGN with \mbh\ derived with the RM and the SE methods. This figure reveals a very good agreement (in terms of intercept, slope, and scatter) with our work, which is based on an IR-selected sample of AGN comprising both obscured and unobscured AGN. Once again, the substantial scatter around the correlation limits its usefulness: no reliable \mbh\ estimates can be obtained from the galaxy stellar mass. Nevertheless this type of correlation plays a crucial role in the study of the evolution of SMBHs and their host galaxies over cosmic time.

%%_____________________________________________________________

%%%%%%%%%%%%%%%%%%%%%%%%%%%%%%%%%%%%%%%%%%
%%%% SECTION 5: DISCUSSION 
%%%%%%%%%%%%%%%%%%%%%%%%%%%%%%%%%%%%%%%%%%
\section{Discussion}
\label{sec5:Discussion}
Because of their crucial importance in our understanding of the cosmic growth of the SMBH within its host galaxy, BH--galaxy correlations have been a very active area of research for years. Historically, the first studies focused on nearby inactive galaxies because only in these objects was the BH sphere of influence spatially resolved and the \mbh\ and  galaxy properties reliably constrained. These investigations revealed the existence of tight correlations suggesting a possible coevolution of BH and host galaxy (e.g., \citealt{Ferrarese2000, Gebhardt2000,Gueltekin2009,Kormendy2013,McConnell2013}). 

Over the years, the advance of observational techniques to determine \mbh\ in AGN and the improvement in SED decomposition separating galaxy and AGN contributions made it possible to extend similar correlations to relatively distant active galaxies. These studies mainly focus on type 1 AGN and can be divided into two main categories: 1) investigations focused on fairly limited samples, where the \mbh\ values are reasonably well constrained using the RM technique (e.g., \citealt{Nelson2004,Woo2010,Woo2013,Winkel2025}) and 2) studies based on larger samples more representative of the AGN population, where the \mbh\ values are estimated with the SE method. The latter technique is characterized by larger uncertainties stemming from scatter around the $R_\mathrm{BLR}$--$L$ luminosity correlation, as well as its extrapolation to untested luminosity and distance domains (e.g., \citealt{Grupe2004,Greene2006,Shen2008,Bennert2011,Caglar2023}).

Unlike previous studies that were restricted to ``well-behaved'' type 1 AGN, we based our work on the 12\,$\mu$m AGN sample, a complete flux-limited, mid-IR selected sample relatively unbiased %with respect to obscuration 
comprising both unobscured (Seyfert 1) and obscured (Seyfert 2 and HBLR) objects, %which is likely to be 
more representative of the local AGN population than optically or X-ray selected samples.

%%%%%%%%%%%%%%%%%%%%%%%%%%%%%%%%%%%%%%%%%%
%%%%%% SEC 5.1 MBH ESTIMATES 
%%%%%%%%%%%%%%%%%%%%%%%%%%%%%%%%%%%%%%%%%%
\subsection{\mbh\ estimates}
 We took advantage of the good quality of \nustar\ data of a sizable portion of the 12\,$\mu$m AGN sample to estimate \mbh\ with the X-ray scaling method, which has been shown to yield reliable results for any AGN accreting at moderate or high level regardless of the level of obscuration.

This method has already been successfully tested on an RM AGN sample \citep{Gliozzi2011}, as well as on heavily obscured AGN with disk maser measurements \citep{Gliozzi2021}, and its reliability has been confirmed and compared with other indirect methods using a volume-limited AGN sample from BASS \citep{Gliozzi2024}. Nevertheless, we carried out a systematic comparison using a subsample of the 12\,$\mu$m AGN with \mbh\ determined with dynamical methods. 

Our analysis validates our previous conclusions: the linear correlation found in Section \ref{sec3.2:mbh} between $M_\mathrm{BH,dyn}$ and $M_\mathrm{BH,X}$ is consistent with the one-to-one correlation. Additionally, the vast majority of $\Delta\log(M_\mathrm{BH})/\sigma_\mathrm{diff}$ values are well below 3 (the average value is 1.5 for the whole sample), as shown in Table \ref{tab:BH_masses}. In other words, we have demonstrated that the X-ray scaling method yields \mbh\ values consistent with the ones determined via dynamical methods regardless of the level of obscuration and accretion, provided that the latter is above a minimum threshold (i.e., Eddington ratio $\gtrsim 10^{-2}$; see, e.g., \citealt{Jang2014}). We therefore conclude that the findings of this work are unlikely to be affected by the use of this specific method to constrain \mbh.

%%%%%%%%%%%%%%%%%%%%%%%%%%%%%%%%%%%%%%%%%%
%%%%%% SEC 5.2 MBH-lambdaEdd 
%%%%%%%%%%%%%%%%%%%%%%%%%%%%%%%%%%%%%%%%%%
\subsection{\mbh--\laedd\ diagram}\label{sec52:Mbh_lambdaEdd}
Having constrained both \mbh\ and the bolometric luminosity values, determined from the direct integration of broadband SEDs obtained with high-resolution observations, we compared the intrinsic properties of the three AGN subclasses (Seyfert 1s, Seyfert 2s, and HBLRs) that are part of the 12\,$\mu$m AGN sample. Based on their level of obscuration from X-ray studies, Seyfert 2s and HBLRs appear to be indistinguishable; however, IR-based investigations reveal that several observational properties of HBLRs appear to be consistent with Seyfert 1s and inconsistent with Seyfert 2s \citep[see, e.g.,][]{Schmitt1998,Tran2001,Tran2003,Tommasin2010,Fernandez2021}. 

Our analysis provides suggestive evidence (at the 2.6$\sigma$ level) that HBLRs have on average \laedd\ values larger than Seyfert 2s (which is likely a heterogeneous class containing some HBLRs together with other nuclei that genuinely lack broad lines), whereas they are consistent with Seyfert 1s. This is in agreement with the findings of \citet{Spinoglio2022}; however, in that work, some \mbh\ values are estimated using the \msigma\ correlation for inactive galaxies, which we demonstrate here to be inconsistent with this AGN sample. The difference in accretion rate between HBLR and ``true'' type 2 AGN (i.e., AGN that lack a BLR) appears to be in agreement with the model proposed by \citet{Nicastro2000},  which hypothesizes that the BLR is formed only if the accretion rate is above a specific threshold. This suggests that HBLRs and genuine type 2 AGN may represent different evolutionary stages (e.g., \citealt{Hunt1999}).

%%%%%%%%%%%%%%%%%%%%%%%%%%%%%%%%%%%%%%%%%%
%%%%%% SEC 5.3 MBH-sigma CORRELATION 
%%%%%%%%%%%%%%%%%%%%%%%%%%%%%%%%%%%%%%%%%%
\subsection{\mbh--galaxy correlations}\label{sec5.3:Mbh_sigma}
The primary goal of this work is the investigation of the \msigma\ correlation of the 12\,$\mu$m AGN sample, whose host galaxies are dominated by spirals. This study was carried out using the most accurate measurements of the bulge velocity dispersion available in the literature: about half of the objects have high-resolution measurements, and for the remaining sources we used the Hyperleda catalog (\citealt{paturel2003}; see Table \ref{tab:general_properties} for details). This correlation is illustrated in Fig.~\ref{fig-msigma} and can be described as follows. The 12\,$\mu$m AGN show a statistically significant positive correlation without apparent differences among the subclasses of Seyfert 1s, Seyfert 2s, and HBLRs. There is substantial scatter (intrinsic scatter $\epsilon=0.45\pm0.10$ dex and rms = 0.68 dex) around a tentative correlation, described by $\log(M_\mathrm{BH}/M_\odot)=(7.50\pm0.17)+(2.66\pm0.77)\times \log(\sigma_\star/200\,\mathrm{km~s^{-1}})$. This is markedly different from the standard \msigma\ correlation obtained for classical bulges in nearby galaxies by \citet{Kormendy2013}, shown in Fig.~\ref{fig-msigma}.

To further explore the reason why many of our AGN have a position in the \msigma\ diagram that is not consistent with the usual relation of \citet{Kormendy2013}, we considered the empirical model proposed by \citet{Chen2020}. The latter shows the scaling laws for $M_{\rm BH}$ as a function of  the stellar-mass surface density $\Sigma_1$ within a projected radius of 1 kpc. 
The evolution proposed in this model moves galaxies from the star-forming ridgeline of the diagram (blue cloud) to the quenched ridgeline (red sequence) by increasing both the black hole mass and the relative mass density, which is directly related to the stellar velocity dispersion.
 
\begin{figure}[ht!]
   \centering
   \includegraphics[width=\hsize]{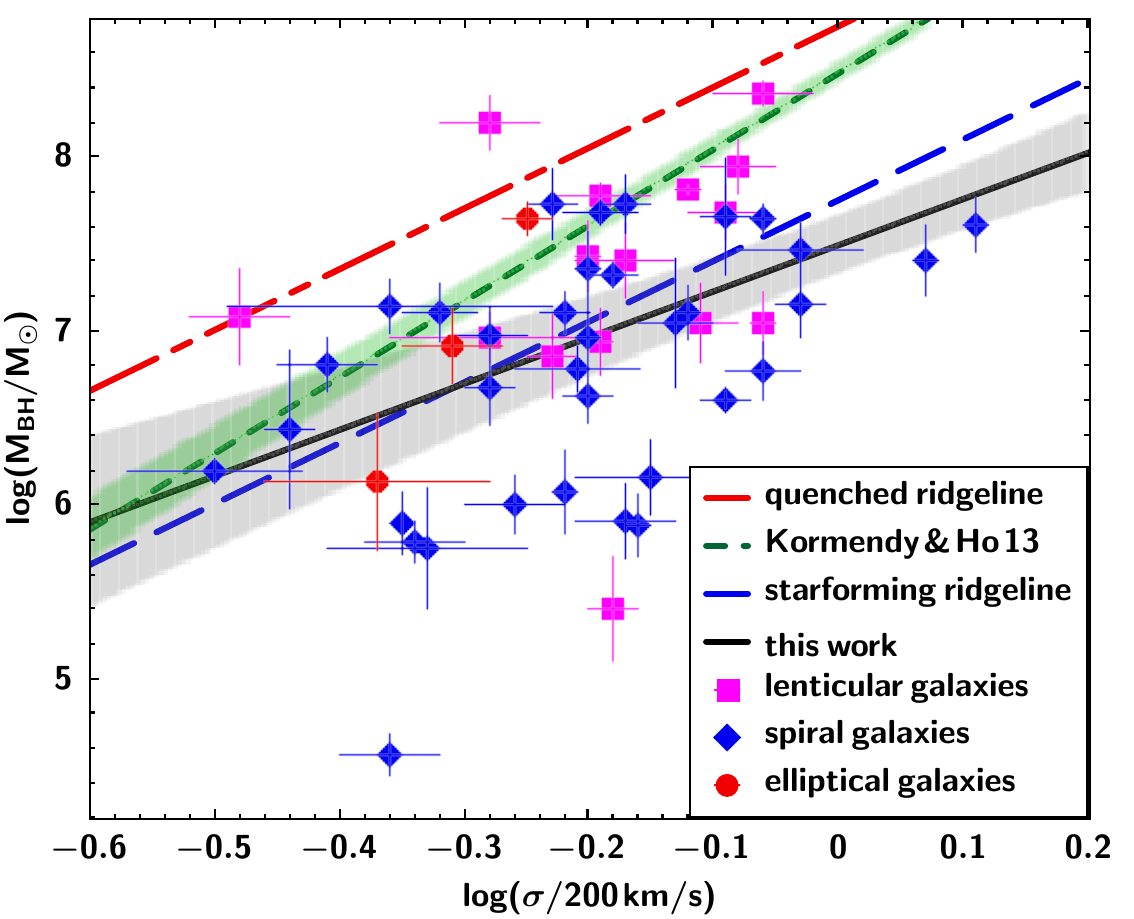}%
      \caption{$\log$(\mbh) plotted versus $\log(\sigma_\star)$ showing the morphological classification of the host galaxies: blue diamonds represent spiral galaxies, magenta squares are lenticular galaxies, and red circles are elliptical galaxies. The black continuous line represents the linear best fit, the green short-dashed line indicates the best fit found by \citet{Kormendy2013} using inactive galaxies with classical bulges, the red long-dashed line indicates the quenched ridgeline and the blue dashed line the star-forming ridgeline from \citet{Chen2020}. The shaded regions account for the errors on the intercepts and slopes of the best-fit linear models.}
         \label{fig-msigma_morpho}
   \end{figure}

In Fig.~\ref{fig-msigma_morpho} we include both the star-forming  and the quenched ridgelines derived from the \mbh\ vs. $\Sigma_1$ diagram \citep[see Fig.~4 of][]{Chen2020}, where  we used the relation of the stellar mass density $\Sigma_1$ to the stellar velocity dispersion $\sigma_{\star}$ of \citet{Fang2013}, valid for the galaxies from the Sloan Digital Sky Survey (SDSS).  It appears from Fig.~\ref{fig-msigma_morpho} that the fit to our AGN population accounting for the uncertainties of the linear fit parameters is broadly consistent with the star-forming ridgeline. This suggests that the AGN in the 12\,$\mu$m sample are still in the black hole growing phase, i.e., in an earlier evolutionary stage compared to more evolved galaxies joining the red sequence.

%_____________________________________________________________________________________
%                        figure 8 log(MBH) vs log(sigma/200km/s) with morphological classification
%-----------------------------------------------------------------------------------------
%----------------------------------------------------------------------------------------------------------- 

To test whether galaxy morphology plays a role in the \msigma\ correlation of the \nustar\ 12\,$\mu$m sample, in Fig.~\ref{fig-msigma_morpho} we plot \mbh\ vs. $\sigma_{\star}$ illustrating the different morphological classifications of the host galaxies taken from the NASA Extragalactic Database\footnote{\url{https://ned.ipac.caltech.edu}}. The sample contains 36 spiral galaxies, 15 lenticular galaxies,  and only 3 elliptical galaxies. A visual inspection of this figure indicates that lenticular galaxies cluster toward the upper part of the plot, close to the linear correlation of inactive galaxies with classical bulges. In contrast, spiral galaxies show lower \mbh\ values well below those predicted by the correlation based on classical bulges. This is consistent with the findings of \citet{Jiang2011} and \citet{Ho2014} based on the \msigma\ correlation of AGN in galaxies with pseudobulges, which showed different linear fits and relatively large scatter compared to classical bulges.

%__________________________________________________________________________________________________________
%                        figure 9 log(MBH_dyn) vs log(sigma/200) with galaxy morphology and SFR color coding
%----------------------------------------------------------------------------------------------------------- 
    \begin{figure}[ht!]
   \centering
   \includegraphics[width=\hsize]{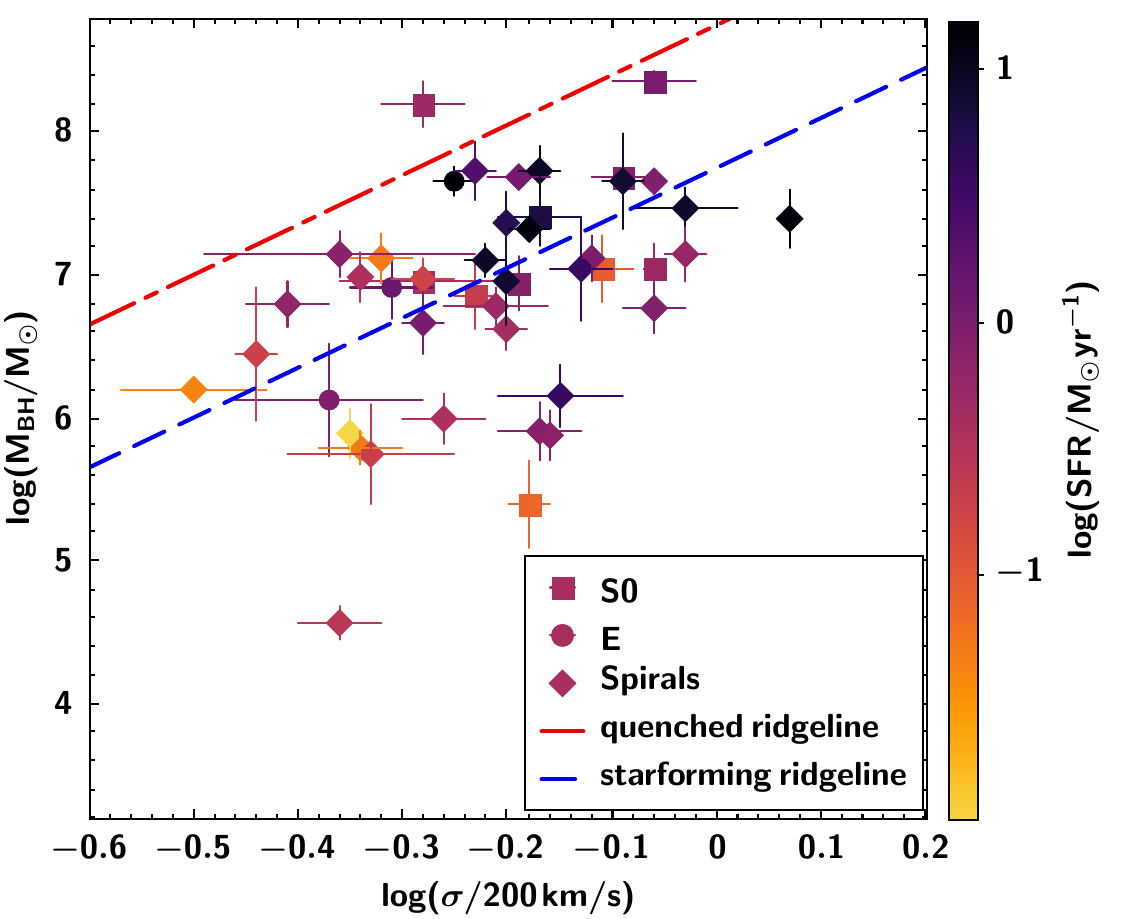}%
      \caption{Same as Fig.~\ref{fig-msigma_morpho}, but color-coding the data points with the value of the star formation rate SFR obtained from the PAH 11.3 $\mu$m line.}
         \label{sSFR}
   \end{figure}
%----------------------------------------------------------------------------------------------------------- 

In Fig.~\ref{sSFR} we investigate if the galaxies located close to or at the bottom of the star-forming ridgeline are characterized by a higher star formation rate (SFR), using the SFR estimates derived from PAH 11.3 $\mu$m  measurements, by color-coding the symbols in the range -2 $<$ log(SFR) $<$ 1. The fluxes of the PAH 11.3 $\mu$m feature are taken from \citet{Spinoglio2022} and the calibration of the SFR from \citet{Xie&Ho2019}. Although S0 and elliptical galaxies appear to be closer to the quenched ridgeline with respect to spiral galaxies, which can be better fitted by the star-forming ridgeline, no clear separation in terms of star formation rate (SFR) can be inferred from this sample due to the large scatter. 
%(in this figure we maintained the same symbols as in Fig.~\ref{fig-msigma_morpho}). 
For completeness, we carried out K--S and Student's $t$ tests to check whether there is any statistically significant difference in the distributions of log(\mbh), log($\sigma_{\star}$), and log(SFR) for the S0 and ellipticals compared to the spiral galaxies. However, no significant differences are found.

Interestingly, in a recent work, \citet{Graham2026} finds that a sample of inactive dust-poor S0 galaxies with \veldis\ in the same range as the 12\,$\mu$m AGN sample and \mbh\ estimated with dynamical measurements shows a significant scatter in the \msigma\ correlation and a flat slope ($\sim 2.5$). This is in agreement with our results. Additionally, previous works based on different samples find lower intercept values and shallower slopes than the \citet{Kormendy2013} correlation. This is the case of \citet{Caglar2023}, using a large volume-limited sample of 154 type 1 AGN from the 105-month BAT sample, which represents the most complete analysis of the \msigma\ correlation of type 1 AGN currently in the literature. A similar result is found using RM AGN samples \citep{Woo2013,Winkel2025} and changing-look AGN \citep{Zeltyn2026}. 

The direct implication of the clear discrepancy between the \msigma\ correlation in the 12\,$\mu$m AGN sample and that obtained from inactive galaxies with classical bulges is that the latter cannot be straightforwardly applied to AGN. As a consequence, the \mbh\ values directly inferred from the \msigma\ correlation obtained using inactive galaxies must be taken with caution since they are likely unreliable (overestimated), as already suggested by previous work (e.g., \citealt{Greene2016,Ricci2017b,Gliozzi2024}). Moreover, the large uncertainties associated with the parameters of the linear fit of the 12\,$\mu$m AGN sample and the substantial scatter suggest that its predictive power in terms of \mbh\ estimates is fairly limited.

Correlations between black hole mass and galaxy bulge play a crucial role in investigating the \mbh\ growth over cosmic time. However, since accurate measurements of galaxy bulge properties are not feasible for distant galaxies because the stellar light is too faint (making it impossible to extend the \msigma\ correlation to high redshifts), we also investigated the correlation between \mbh\ and the galaxy stellar mass $M_\star$, which is illustrated in Fig.~\ref{fig-MM*}. 
As already anticipated in Sect. \ref{sec4.3:MMgal}, our best-fit model is nearly indistinguishable from the one derived by \citet{Reines2015} using a large sample of broad-line AGN with \mbh\ obtained with RM and the SE method, 
which, in turn, is shown to be inconsistent with the correlations obtained with inactive galaxies (e.g., \citealt{Haring2004,Kormendy2013,McConnell2013}). Massive host galaxies are also characteristic of the local AGN population more generally. For example, \citet{Kauffmann2003}, using a large SDSS sample, showed that local AGN preferentially reside in massive galaxies, with the most powerful systems associated with hosts that show evidence for younger stellar populations. Our results extend this demographic picture by examining how the black hole mass itself scales with the total stellar mass in a mid-infrared-selected AGN sample.

Recent JWST \citep{gardner2006} observations have revealed substantial populations of actively accreting black holes at high redshift, including systems with moderate AGN luminosities as well as much more luminous objects (e.g., \citealt{Larson2023,Matthee2024,Greene2024,Maiolino2024,Lambrides2024,Lupi2024}). Our results do not imply that these distant AGN are direct analogs of the local 12 micron sample. Rather, they emphasize more generally that black hole–galaxy relations calibrated from local inactive galaxies should be applied cautiously to actively accreting systems, particularly when the redshift, luminosity, and host-galaxy properties differ substantially from those of the calibration samples.

%%%%%%%%%%%%%%%%%%%%%%%%%%%%%%%%%%%%%%%%%%
%%%% SECTION 6: CONCLUSIONS
%%%%%%%%%%%%%%%%%%%%%%%%%%%%%%%%%%%%%%%%%%
\section{Conclusions}
Our main results can be summarized as follows:
   \begin{enumerate}
      \item We confirm that the X-ray scaling method yields estimates of \mbh\ (with typical uncertainties of the order of 0.1--0.3 dex) that are consistent with dynamical values for AGN with and without detected broad lines (Sect. \ref{sec3.2:mbh}). 
      \item We compare the intrinsic properties of HBLRs, Seyfert 1s, and Seyfert 2s (the three subclasses of the 12\,$\mu$m AGN sample) and find marginal evidence that the \laedd\ distribution of HBLRs is consistent with Seyfert 1s and has larger values of \laedd\ than Seyfert 2s (Sect. \ref{sec4.1:mbhlaedd}). This suggests that the two optically obscured subclasses of Seyfert galaxies (HBLRs and Seyfert 2) are intrinsically different.
      \item We find that the \msigma\ correlation of the 12\,$\mu$m AGN sample is characterized by a substantial scatter and appears to be markedly different from the one obtained with nearby inactive galaxies with classical bulges (Sect. \ref{sec4.2:Msigma}).  Specifically, the significantly lower black hole masses for a given stellar velocity dispersion of our sample of AGN can be interpreted as an earlier evolutionary stage of these galaxies, which are supposedly evolving from the blue cloud to the red sequence, accreting and increasing their M$_{\rm BH}$ and possibly still forming stars (Sect. \ref{sec5.3:Mbh_sigma}). The different \msigma\ correlation with respect to inactive galaxies implies that the latter should be used with caution to estimate the \mbh\ in AGN where the BLR is not accessible and that the \msigma\ correlation may have a limited predictive power for \mbh\ estimates.
      \item We find that the \mbh--$M_\star$ correlation based on the 12\,$\mu$m AGN sample appears to be fully consistent with the analogous correlation obtained from a large sample of nearby type 1 AGN and inconsistent with correlations based on inactive galaxies (Sect. \ref{sec4.3:MMgal}). This reinforces the conclusion that BH--galaxy correlations obtained using inactive galaxies are different from the ones using AGN and should be used with caution to derive any AGN property. 
   
   \end{enumerate}

\noindent \textbf{Data availability}\\
\noindent All the data used in this work are publicly available in the archives.
Tables B.1, B.2 and B.3 are only available in electronic form at the CDS via anonymous ftp to cdsarc.u-strasbg.fr (130.79.128.5) or via http://cdsweb.u-strasbg.fr/cgi-bin/qcat?J/A+A/.

%%%%%%%%%%%%%%%%%%%%%%%%%%%%%%%%%%%%%%%%%%%%%%%%%%%%%%%%%%%%%%
\begin{acknowledgements}
      We thank the anonymous referee for the useful comments that improved the clarity of the paper.
      This research has made use of the NASA/IPAC Extragalactic Database (NED), which is funded by the National Aeronautics and Space Administration and operated by the California Institute of Technology. 
      JAFO acknowledges financial support by the Spanish Ministry of Science, Innovation and Universities (MICIU/AEI/10.13039/501100011033), by ``ERDF A way of making Europe'' and by ``European Union NextGenerationEU/PRTR'' through the grants PID2021-124918NB-C44 and CNS2023-145339; MCIN and the European Union -- NextGenerationEU through the Recovery and Resilience Facility project ICTS-MRR-2021-03-CEFCA; MICIU and ERDF/EU through the grant PID2023-147386NB-I00. 
\end{acknowledgements}

% Please note that we have included the references below in
% order to compile the document, but we ask you to:

 \bibliographystyle{aa} % style aa.bst 
 \bibliography{BHgal12mu} % your references Yourfile.bib

%%%%%%%%%%%%%%%%%%%%%%%%%%%%%%%%%%%%%%%%%%%%%%%%%%%%%%%%%%%%%%%
% Appendices must be placed after   \end{thebibliography}
% They will be placed automatically on a new page.
%%%%%%%%%%%%%%%%%%%%%%%%%%%%%%%%%%%%%%%%%%%%%%%%%%%%%%%%%%%%%%%
\begin{appendix}
%%%%%%%%%%%%%%%%%%%%%%%%%%%%%%%%%%%%%%%%%%%%%%%%%%%%%%%%%%%%%%%

\section{X-ray scaling method}

This method, whose technical details are discussed at length in previous works \citep{Shaposhnikov2009,Gliozzi2011,Gliozzi2021,Williams2023,Gliozzi2025}, is based on two reasonable assumptions: 1) AGN behave similarly to stellar-mass BHs in X-ray binaries: specifically, the mechanism producing hard X-ray emission in all accreting BH systems is universal (e.g., \citealt{Done2005,Koerding2006,McHardy2006}) and 2) the hard X-ray photon index is a reliable indicator of the accretion state of the system, as supported by several studies based on stellar-mass and supermassive BHs in both unabsorbed and absorbed AGN (e.g., \citealt{Remillard2006,Shemmer2008,Sobolewska2009,Risaliti2009,Brightman2013,Brightman2016,Serafinelli2017}). 

Here we summarize the basic steps to compute \mbh\ with this method. In short, the mass of SMBHs in AGN is obtained by scaling up the dynamically constrained \mbh\ of well-studied X-ray binaries, our reference sources. As in our previous works, the stellar-mass reference sources used here are GRO J1655-40, GX 339-4, and XTE J1550-564, whose dynamical \mbh\ values are constrained with an uncertainty of $15\%$ or better (mass and distances with their relative uncertainties, as well as spectral trend best-fit parameters, are reported in \citealt{Shaposhnikov2009,Gliozzi2011,Williams2023}). 

The scaling factor of this method is the ratio $L_\mathrm{AGN}/L_\mathrm{ref}$ of the AGN's to the reference source's luminosities. Both luminosities are proportional to their respective \mbh, accretion rate in Eddington units $\dot m$, and radiative efficiency $\eta$ values. 
In our analysis we obtain the accretion luminosity values from the normalization of the bulk motion Comptonization model \texttt{BMC} \nbmc\ (which is in units of luminosity divided by the distance squared). This is a simple Comptonization model that can describe both bulk motion and thermal Comptonization and can be applied to X-ray spectra from AGN as well as from stellar-mass BH systems \citep{Titarchuk1997}. In mathematical terms, our scaling factor is
\[
 \left(\frac{L_{\mathrm{AGN}}}{L_{\mathrm{ref}}}\right) =
\left(\frac{N_{\mathrm{BMC,AGN}}}{N_{\mathrm{BMC,ref}}}\right) 
 \times \left(\frac{d_{\mathrm{AGN}}^2}{d_{\mathrm{ref}}^2}\right)
\]

While for the AGN we have a single value of $L$, for each reference source we have several values, since actively accreting stellar-mass BHs continuously transition between different spectral states over timescales of several weeks or a few months. The assumption that the photon index $\Gamma$ is a reliable indicator of the accretion state is what makes it possible to compare the AGN luminosity to the reference source luminosity in the appropriate  accretion state. Importantly, regardless of their mass, BH systems in the same accretion state are characterized by the same values of $\dot m$ and  $\eta$. This implies that  $L_\mathrm{AGN}/L_\mathrm{ref}=M_\mathrm{BH, AGN}/M_\mathrm{BH,ref}$ and leads to the equation used to determine the AGN \mbh\ value:
\[
M_{\mathrm{BH,AGN}}=M_{\mathrm{BH,ref}}  \times \left(\frac{N_{\mathrm{BMC,AGN}}}{N_{\mathrm{BMC,ref}}}\right)  \times \left(\frac{d_{\mathrm{AGN}}^2}{d_{\mathrm{ref}}^2}\right)
\]

\begin{figure}[ht!]
   \centering
   \includegraphics[width=\hsize]{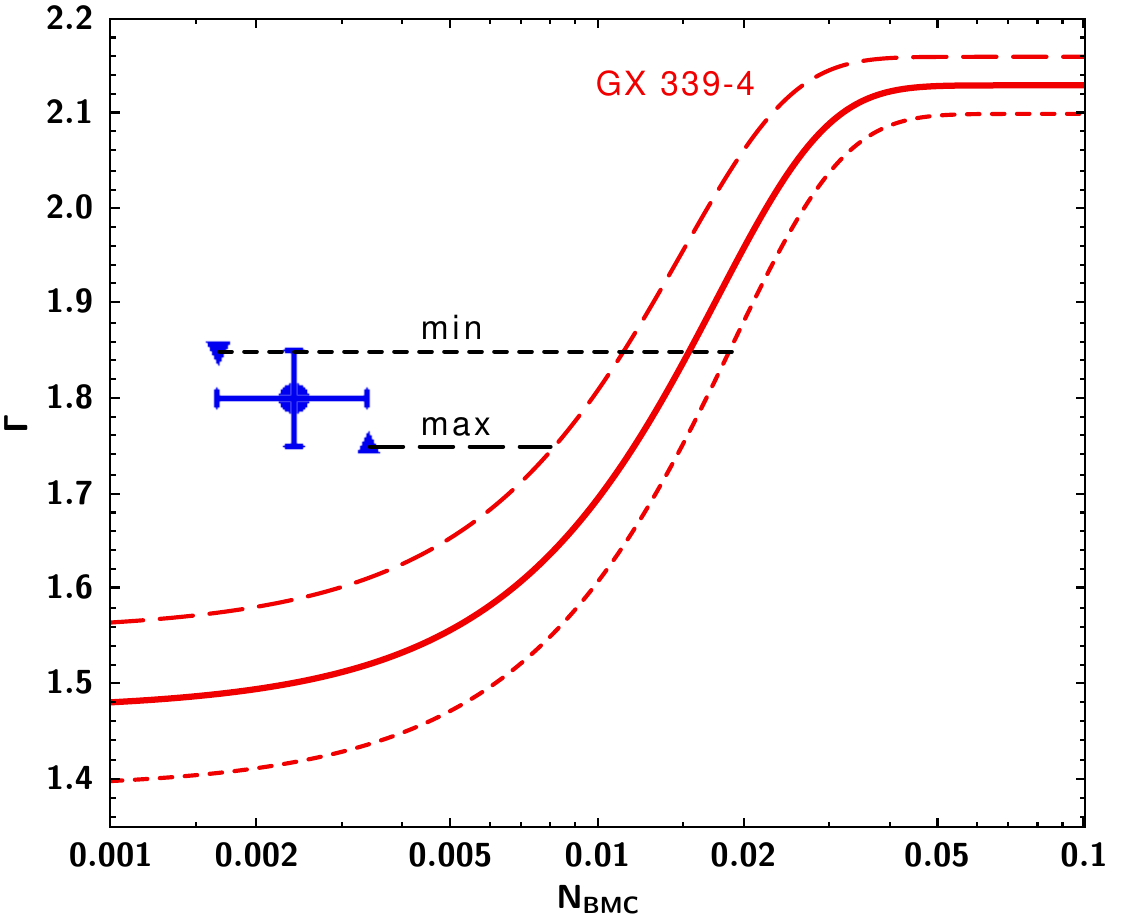}%{scal1.pdf} 
      \caption{\nbmc\ -- $\Gamma$ diagram illustrating the position of the AGN (blue circle) with its uncertainties on \nbmc\ and $\Gamma$ and the spectral trend of the reference source \gx\ during the 2003 decay (continuous red line) and its 1$\sigma$ uncertainties (long- and short-dashed red lines). The upper limit on \mbh\ is obtained by the separation between the point ($N_\mathrm{BMC} + 1\sigma$, $\Gamma - 1\sigma$), represented by the upward-pointing blue triangle, and the upper envelope of the spectral trend (long-dashed red line). This separation is shown by the long-dashed black line labeled max. On the other hand, the lower limit on \mbh\ is given by the horizontal separation between the point ($N_\mathrm{BMC} -1\sigma$, $\Gamma + 1\sigma$), represented by the downward-pointing blue triangle, and the lower envelope of the spectral trend (short-dashed red line). This separation is shown by the short-dashed black line labeled min.
      }
         \label{figscal1}
   \end{figure}

This scaling process is illustrated in Fig.~\ref{figscal1}, which reports the position of the AGN and the spectral trend of the stellar-mass reference source \gx\ during the 2003 transition from the high-soft state to the low-hard spectral state in the \nbmc--$\Gamma$ diagram. First, the value of $\Gamma$ (that is, the y-value of the AGN in the \nbmc--$\Gamma$ diagram) maps the AGN to the corresponding accretion state of the reference source. Then, the separation between the AGN and the reference \nbmc\ values allows one to scale \mbh, once the respective distances are taken into account. The smaller the separation in \nbmc\ values between the AGN and reference source, the larger the value of $M_\mathrm{BH,AGN}$.
%__________________________________________________________________________________________________________
%                        figure A1 N_BMC-Gamma plane with AGN and GX reference source
%----------------------------------------------------------------------------------------------------------- 
%----------------------------------------------------------------------------------------------------------- 

Fig. \ref{figscal1} also illustrates how the \mbh\ uncertainties are estimated.  Depending on the quality of the AGN X-ray spectrum and the reference source utilized, the uncertainties of $\log(M_\mathrm{BH,X})$ typically range between 0.1 and 0.3 dex.
%__________________________________________________________________________________________________________
%                        figure A2 N_BMC-Gamma plane with AGN and GX reference source
%----------------------------------------------------------------------------------------------------------- 
    \begin{figure}[ht!]
   \centering
   \includegraphics[width=\hsize]{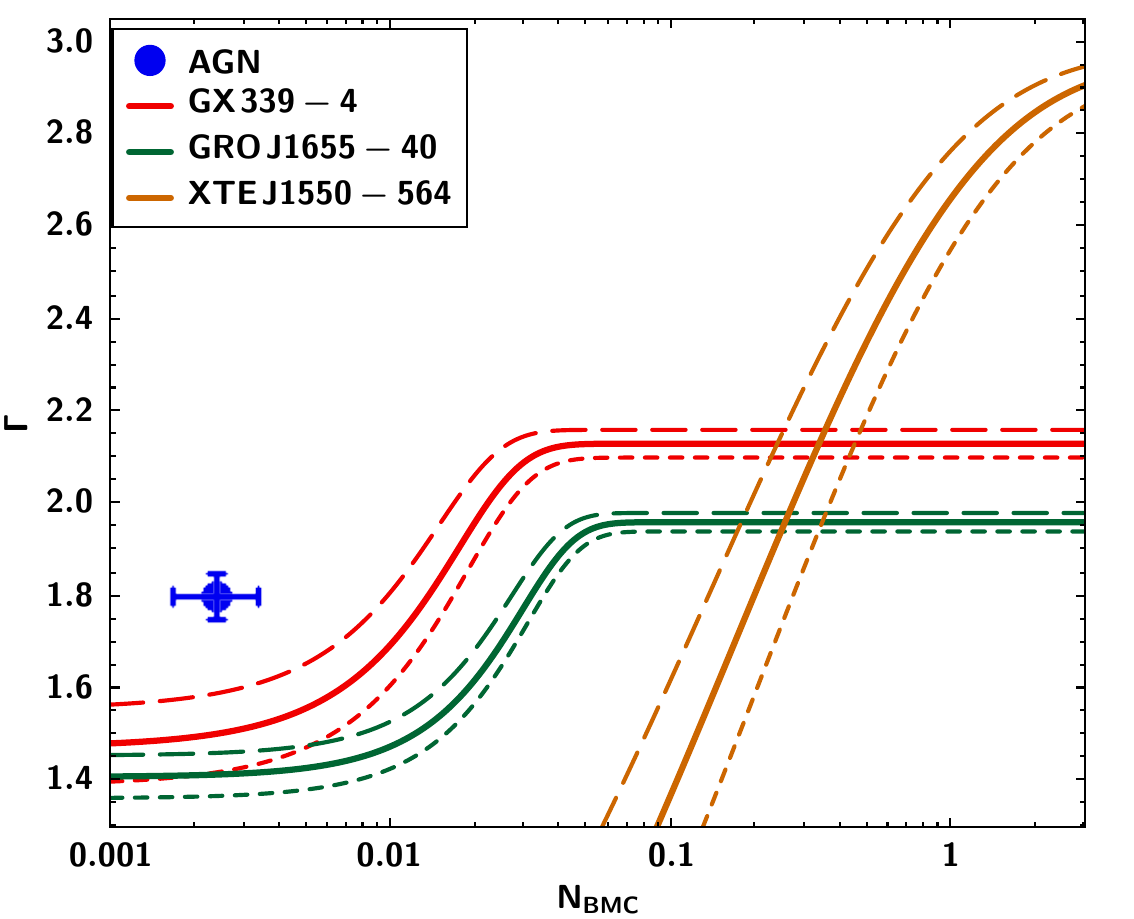}%{scal2.pdf} 
      \caption{\nbmc\ -- $\Gamma$ diagram with the three most reliable reference sources: \gx\ during the 2003 decay (red), \gro\ during the 2005 decay (green), and \xte\ during the 1998 rise (brown). The continuous lines indicate the best-fit spectral trends, whereas the long-dashed and short-dashed lines indicate their 1$\sigma$ upper and lower uncertainties, respectively.}
         \label{figscal2}
   \end{figure}
%----------------------------------------------------------------------------------------------------------- 

As illustrated in Figure \ref{figscal2}, different reference sources span different ranges of $\Gamma$ during their spectral transitions; therefore, we use the most appropriate reference source depending on the AGN photon index value. More specifically, for AGN with $\Gamma$ in the 1.6--1.9 range, both \gro\ and \gx\ yield \mbh\ values consistent within their respective uncertainties and hence we use the average value obtained with these two reference sources. For uncertainties, to be conservative, we use the \gx\ upper limit as maximum and the \gro\ lower limit as \mbh\ minimum. For $1.5 <\Gamma <1.6$, only GRO J1655-40 yields reliable scaled \mbh\ values, whereas GX 339-4 is used for $1.9 <\Gamma <2.1$. Finally, for very steep $\Gamma$ values, we utilize XTE J1550-564 as a reference with a multiplicative factor of 4, which was empirically determined in previous work \citep{Williams2023,Gliozzi2024}.

Whereas the X-ray scaling method provides a valid alternative to optically based methods to determine the \mbh\ in AGN, it has some specific limitations. First, it requires good quality X-ray spectra, which means with sufficient S/N to tightly constrain the spectral parameters. More specifically, for \nustar\ observations this is achieved with spectra containing at least several hundred counts. The median number of counts of the spectra used in this work is 8600, with a minimum of $\sim$800 counts in NGC 3982 and Mrk 897. This requirement severely limits the applicability of this method compared to optical studies. Second, the X-ray scaling method is limited to BH systems accreting at or above the 1\% threshold in Eddington units. Below this threshold the softer-when-brighter trend (which is the basis for the X-ray scaling method) disappears, and at even lower values of $\dot{m}$, an inverse correlation between $\Gamma$ and luminosity is observed (e.g, \citealt{Gu2009,Constantin2009,Jang2014}).

%____________________________________________________________
% A long table in appendix
%------------------------------------------------------------
% This is the start of the page
\onecolumn
\section{Characteristics of the 12\,$\mu$m sample}

\begin{longtable}{lllrrrl}
\caption{General properties} \\
\label{tab:general_properties} \\

\toprule
\midrule
Source & Type & Class & $\log \left(\frac{L_\mathrm{bol}}{\mathrm{erg\; s^{-1}}}\right)$  & $\log\left(\frac{M_\star}{M_\odot}\right)$ & $\sigma_\star\,({\mathrm{km\,s^{-1}}})$ & Reference  \\
(1) & (2) & (3) & (4) & (5) & (6) & (7) \\
\midrule
\endfirsthead

\caption{continued.} \\
\toprule
\midrule
Source & Type & Class & $\log \left(\frac{L_\mathrm{bol}}{\mathrm{erg\; s^{-1}}}\right)$  & $\log\left(\frac{M_\star}{M_\odot}\right)$ & $\sigma_\star\,({\mathrm{km\,s^{-1}}})$ & Reference  \\
(1) & (2) & (3) & (4) & (5) & (6) & (7) \\
\midrule
\endhead

\bottomrule
\endfoot

\bottomrule
\endlastfoot
Mrk 335 & S0-a & Sy1 & 44.8 & 10.1 & $66\pm6$ & \citet{Winkel2025} \\
Mrk 348 & S0-a & HBLR & 44.5 & 9.72 & $104.7\pm9.4$ & Hyperleda \\
UGC 545 & Sc   & Sy1 & 45.7 & 10.77 &  $186\pm22$ & \citet{Molina2024} \\
NGC 424 & S0-a & HBLR & 44.4 & 9.84 & $118\pm5$ & \citet{Koss2022} \\
NGC 513 & Sc   &  HBLR & 43.9 & 10.03 & $148.3\pm9.8$ & Hyperleda \\
IRAS F01475-0740 & E-S0 & HBLR & 44.4 & 9.06 & $84.4\pm17.9$ & Hyperleda \\
NGC 931 & Sbc  & Sy1 & 44.5 & 10.8 & $130\pm6$ & \citet{Caglar2023} \\
NGC 1068 & Sb  & HBLR & 44.5 & 10.24 & $161.8\pm6.8$ & Hyperleda \\
NGC 1125 & S0-a & HBLR & 43.6 & $\dots$ & $131.9\pm7.6$ & Hyperleda \\
NGC 1144 & S pec & Sy2 & 45.0 & $\dots$ & $101\pm3$ & \citet{Koss2022} \\
MCG -02-08-039 & Sa & HBLR & 44.2 & 10.16 & $\dots$ & $\dots$ \\
NGC 1194 & S0-a & Sy1 & 44.4 & 9.78 & $151\pm5$ & \citet{Koss2022} \\
NGC 1320 & Sa & Sy2 & 43.8 & 9.57 & $109.6\pm10.4$ & Hyperleda \\
NGC 1365 & Sb & Sy1 & 43.8 & 10.35 & $141.1\pm18.7$ & Hyperleda \\
NGC 1386 & S0-a & HBLR & 43.5 & 9.12 & $133.1\pm6.7$ & Hyperleda \\
IRAS F03362-1642 & SBb & Sy2 & $> 43.8$ & $\dots$ & $\dots$ & $\dots$ \\
NGC 1566 & SABb & Sy1 & 42.8 & 10.07 & $135\pm11$ & \citet{Koss2017} \\
3C 120 & S0 & Sy1 & 45.4 & 9.86 & $172.6\pm17.2$ & Hyperleda \\
Mrk 618 & SBb & Sy1 & 44.8 & 10.29 & $134.4\pm38.8$ & \citet{Bosch2015} \\
IRAS F04385-0828 & S0-a & HBLR & 43.9 & 9.34 & $\dots$ & $\dots$ \\
ESO 033-G002 & S0 & Sy2 & 44.1 & 9.9 & $125\pm3$ & \citet{Koss2022} \\
ESO 362-G018 & S0-a & Sy1 & 43.9 & 10.24 & $129.9\pm4.2$ & Hyperleda \\
IRAS F05189-2524 & E & HBLR & 45.4 & 10.38 & $113\pm4$ & \citet{Koss2022} \\
ESO 253-G003 & Sa & Sy2 & $\dots$ & 10.14 & $134\pm5$ & \citet{Caglar2023} \\
Mrk 6 & S0-a & Sy1 &  44.5 & 10.06 & $\dots$ & $\dots$ \\
Mrk 9 & S0 & Sy1 & 44.7 & 10.28 & $136\pm11$ & \citet{Caglar2023} \\
Mrk 79 & Sb & Sy1 & 44.9 & 10.31 & $129.5\pm10$ & Hyperleda \\
Mrk 704 & S0-a & Sy1 & 44.9 & 10.52 & $282.6\pm20.3$ & \citet{Bosch2015} \\
NGC 2992 & Sa & Sy1 & 44.1 & 9.75 & $152\pm4$ & \citet{Koss2022} \\
Mrk 1239 & E & Sy1 & 44.6 & 10.07 & $99\pm8$ & \citet{Winkel2025} \\
3C 234 & N & Sy1 & 46.1 & 10.69 & $\dots$ & $\dots$ \\
NGC 3079 & Sbc & Sy2 & 43.3 & 9.65 & $175\pm12.4$ & Hyperleda \\
NGC 3227 & SABa & Sy1 & 43.3 & 9.06 & $126.8\pm5.8$ & Hyperleda \\
NGC 3516 & S0 & Sy1 & 44.0 & 10.41 & $153.6\pm10.8$ & Hyperleda \\
NGC 3982 & SABb & HBLR & 43.9 & 10.07 & $71.8\pm3.9$ & Hyperleda \\
NGC 4051 & SABb & Sy1 & 42.8 & 8.5 & $90.6\pm9.1$ & Hyperleda \\
UGC 7064 & SBb & HBLR & 43.9 & 10.13 & $127\pm3$ & \citet{Koss2022} \\
NGC 4151 & Sab & Sy1 & 43.9 & 9.37 & $95\pm7$ & \citet{Dasyra2011} \\
Mrk 766 & Sba & Sy1 & 44.3 & 10.23 & $87.1\pm26$ & \citet{Sasseville2025} \\
NGC 4388 & SABb & HBLR & 44.1 & 10.07 & $105\pm6$ & \citet{Koss2022} \\
NGC 4593 & Sb & Sy1 & 44.2 & 10.13 & $122.3\pm13.7$ & Hyperleda \\
NGC 4748 & Sa & Sy1 & 43.8 & 9.8 & $77.2\pm7$ &  Hyperleda \\
Mrk 231 & Sc & Sy1 & $> 45.6$ & 11.04 & $\dots$ & $\dots$ \\
NGC 4941 & SABa & Sy2 & 42.9 & 9.25 & $90\pm1$ & \citet{Koss2022} \\
NGC 5033 & Sc & Sy1 & 42.2 & 9.57 & $139\pm2$ & \citet{Koss2022} \\
MCG -03-34-064 & S0 & HBLR & 44.5 & 9.96 & $176\pm4$ & \citet{Koss2022} \\
NGC 5135 & Sab & Sy2 & 44.4 & 10.16 & $125.5\pm3.7$ & Hyperleda \\
NGC 5194 & SABb & Sy2 & 41.8 & 9.58 & $87.9\pm7.3$ & Hyperleda \\
MCG -06-30-015 & Sab & Sy1 & 44.0 & 9.31 & $105.2\pm7.8$ & Hyperleda \\
IRAS F13349+2438 & $\dots$ & Sy2 & 46.1 & 11.15 & $\dots$ & $\dots$ \\
Mrk 266 & pec & Sy2 & 44.0 & 10.39 & $187\pm8$ & \citet{Dasyra2011}  \\
Mrk 273 & Sab & Sy2 & 44.6 & 10.17 & $235\pm7$ &  \citet{Koss2022}\\
IC 4329A & S0-a & Sy1 & 45.2 & 10.33 & $166\pm12$ & \citet{Winkel2025} \\
NGC 5347 & Sab & HBLR & 43.4 & 9.34 & $92.6\pm16.2$ & Hyperleda \\
Mrk 463 & Sc & HBLR & 45.4 & 10.67 & $163\pm8$ & \citet{Dasyra2011}\\
NGC 5506 & SABa & Sy2 & 44.2 & 9.68 & $104\pm19$ & \citet{Koss2022} \\
NGC 5548 & S0-a & Sy1 & 44.5 & 10.35 & $163\pm13$ & \citet{Winkel2025} \\
Mrk 817 & S0-a & Sy1 & 45.0 & 10.2 & $128.3\pm12.1$ & Hyperleda \\
IRAS F15091-2107 & Sab & Sy1 & 45.2 & 10.36 & $\dots$ & $\dots$ \\
NGC 5929 & Sa & HBLR & 43.1 & $\dots$ & $119.9\pm3$ & Hyperleda \\
MCG -02-40-004 & SABa & HBLR & 44.8 & 11 & $\dots$ & $\dots$ \\
IRAS F15480-0344 & E & HBLR & 44.6 & 10.05 & $\dots$ & $\dots$ \\
ESO 141-G055 & Sb & Sy1 & 45.3 & 10.38 & $\dots$ & $\dots$ \\
NGC 6890 & SBb & Sy2 & 43.6 & 10.76 & $\dots$ & $\dots$ \\
Mrk 509 & $\dots$ & Sy1 & 45.7 & 10.53 & $130\pm10$ & \citet{Winkel2025} \\
IC 5063 & S0-a & HBLR & 44.4 & 10.03 & $175\pm4$ &  \citet{Koss2022} \\
Mrk 897 & E-S0 & Sy2 & $>44.0$ & 9.35 & $\dots$ & $\dots$ \\
NGC 7172 & Sa & Sy2 & 44.0 & 9.97 & $186.2\pm7.7$ & Hyperleda \\
3C 445 & N & Sy1 & 45.4 & 10.17 & $\dots$ & $\dots$ \\
NGC 7314 & SABb & HBLR & 43.4 & 9.51 & $63\pm10$ & \citet{Koss2022} \\
NGC 7469 & Sa & Sy1 & 44.7 & 10.28 & $132.9\pm4.8$ & Hyperleda \\
NGC 7582 & SBab & Sy2 & 43.9 & 9.92 & $118.4\pm5.2$ & Hyperleda \\
%NGC 7590 & Sbc & Sy2 & 42.8 & 9.49 & $91.5\pm2.5$ & Hyperleda \\
NGC 7674 & SBb & HBLR & 45.0 & 10.39 & $121\pm5$ & \citet{Koss2022} \\
\end{longtable}
\flushleft
Columns: 1 = source name, 2 = galaxy morphological type, 3 = AGN class (Sy1 = Seyfert 1, Sy2 = Seyfert 2, HBLR = hidden broad-line region), 4 = bolometric luminosity, as derived from \citet{Spinoglio2024} and rescaled taking into account the different cosmological parameters, 5 = stellar mass of the host galaxy from \citet{Spinoglio2022}, 6 = stellar velocity dispersion, 7 = references for the stellar velocity dispersion, for the Hyperleda catalog: \citet{paturel2003}.
%} % end \longtab

%\pagebreak

%\pagebreak

\longtab[2]{
\begin{longtable}{llcccccc}
\caption{X-ray properties} \\
\label{tab:x-ray_properties} \\

\toprule
\midrule
%\noalign{\smallskip}
Source & \nustar  & Exposure & $\log \left(\frac{N_\mathrm{H,los}}{\mathrm{cm}^{-2}}\right)$ & $\Gamma$ & $N_\mathrm{BMC}$ & $\chi^2_\mathrm{red}$ (dof) & $\log \left(\frac{L_\mathrm{X}}{\mathrm{erg\; s^{-1}}}\right)$ \\
 & obsid & time [ks] & & & & & \\
(1) & (2) & (3) & (4) & (5) & (6) & (7) & (8) \\
\midrule
\endfirsthead

\caption{continued.} \\
\toprule
\midrule
Source & \nustar  & Exposure & $\log \left(\frac{N_\mathrm{H,los}}{\mathrm{cm}^{-2}}\right)$ & $\Gamma$ & $N_\mathrm{BMC}$ & $\chi^2_\mathrm{red}$ (dof) & $\log \left(\frac{L_\mathrm{X}}{\mathrm{erg\; s^{-1}}}\right)$ \\
 & obsid & time [ks] & & & & & \\
(1) & (2) & (3) & (4) & (5) & (6) & (7) & (8) \\
\midrule
\endhead

\bottomrule
\endfoot

\bottomrule
\endlastfoot
%\noalign{\smallskip}
Mrk 335 & 60001041002 & 21 & 22.9 & $1.91 \pm 0.03$ & $5.9^{+0.3}_{-0.3}\times 10^{-5}$ & 1.13 (190) & 43.4 \\
\noalign{\smallskip}
Mrk 348 & 60701017002 & 105 & 23.1 & $1.81 \pm 0.01$ & $1.8^{+0.1}_{-0.1}\times 10^{-3}$ & 1.02 (1633) & 43.3 \\
\noalign{\smallskip}
UGC 545 & 60501030002 & 233 & 20.7 & $2.39 \pm 0.01$ & $7.5^{+0.1}_{-0.1}\times 10^{-5}$ & 1.08 (649) & 43.8 \\
\noalign{\smallskip}
NGC 424 & 60662007002 & 34 & 23.4 & $1.81 \pm 0.02$ & $2.0^{+0.1}_{-0.1}\times 10^{-4}$ & 1.10 (234) & 43.0 \\
\noalign{\smallskip}
NGC 513 & 60001153002 & 16 & 22.7 & $1.43 \pm 0.06$ & $2.0^{+0.1}_{-0.2}\times 10^{-5}$ & 0.90 (80) & 42.7 \\
\noalign{\smallskip}
IRAS F01475-0740 & 60360005002 & 31 & 24.0 & $2.45 \pm 0.13$ & $4.0^{+1.1}_{-1.0}\times 10^{-5}$ & 0.89 (45) & 42.1 \\
\noalign{\smallskip}
NGC 931 & 60101002004 & 64 & 22.3 & $2.01 \pm 0.01$ & $3.0^{+0.1}_{-0.1}\times 10^{-4}$ & 1.00 (1131) & 43.3 \\
\noalign{\smallskip}
NGC 1068 & 60002033002 & 52 & 24.5 & $1.98 \pm 0.01$ & $2.2^{+0.1}_{-0.1}\times 10^{-3}$ & 1.06 (713) & 42.4 \\
\noalign{\smallskip}
NGC 1125 & 60510001002 & 32 & 23.7 & $1.78 \pm 0.03$ & $6.5^{+0.5}_{-0.5}\times 10^{-5}$ & 1.00 (94) & 42.5 \\
\noalign{\smallskip}
NGC 1144 & 60368001002 & 21 & 23.6 & $1.60 \pm 0.03$ & $7.6^{+0.4}_{-0.5}\times 10^{-5}$ & 0.65 (121) & 43.7 \\
\noalign{\smallskip}
MCG -02-08-039 & 60360010002 & 27 & 23.9 & $1.79 \pm 0.14$ & $2.5^{+0.6}_{-0.5}\times 10^{-5}$ & 0.79 (41) & 42.6 \\
\noalign{\smallskip}
NGC 1194      & 60061035002 & 32 & 23.9 & $1.62 \pm 0.05$ & $6.4^{+1.1}_{-0.8}\times 10^{-5}$ & 1.16 (167) & 42.6 \\
\noalign{\smallskip}
NGC 1320 & 60061036004 & 28 & 24.1 & $1.70 \pm 0.07$ & $3.9^{+0.7}_{-0.7}\times 10^{-5}$ & 1.21 (61) & 42.0 \\
\noalign{\smallskip}
NGC 1365 & 60702058002 & 56 & 23.3 & $1.69 \pm 0.01$ & $1.4^{+0.1}_{-0.1}\times 10^{-4}$ & 0.97 (828) & 41.7 \\
\noalign{\smallskip}
NGC 1386 & 60201024002 & 26 & 24.5 & $1.91 \pm 0.09$ & $4.6^{+1.5}_{-1.0}\times 10^{-5}$ & 1.27 (39) & 40.9 \\
\noalign{\smallskip}
IRAS F03362-1642 & 60362030002  & 45 & 23.5 & $1.52 \pm 0.04$ & $1.4^{+0.3}_{-0.2}\times 10^{-5}$ & 1.05 (54) & 41.9 \\
\noalign{\smallskip}
NGC 1566 & 60501031002 & 59 & 20.9 & $1.82 \pm 0.02$ & $1.1^{+0.1}_{-0.1}\times 10^{-4}$ & 0.98 (916) & 41.9 \\
\noalign{\smallskip}
3C 120 & 61002011004 & 47 & 21.5 & $1.98 \pm 0.01$ & $6.1^{+0.1}_{-0.1}\times 10^{-4}$ & 0.99 (1127) & 44.2 \\
\noalign{\smallskip}
Mrk 618 & 60761003002 & 21 & 20.7 & $1.90 \pm 0.01$ & $1.4^{+0.1}_{-0.1}\times 10^{-4}$ & 0.95 (372) & 43.5 \\
\noalign{\smallskip}
IRAS F04385-0828 & 60701043002 & 31 & 24.7 & $1.57 \pm 0.09$ & $2.6^{+0.8}_{-0.5}\times 10^{-5}$ & 0.86 (47) & 42.9 \\
\noalign{\smallskip}
ESO 033-G002 & 60601002002 & 175 & 22.8 & $2.14 \pm 0.02$ & $3.4^{+0.3}_{-0.3}\times 10^{-4}$ & 1.02 (1059) & 42.3 \\
\noalign{\smallskip}
ESO 362-G018 & 60201046002 & 102 & 22.1 & $1.75 \pm 0.01$ & $1.1^{+0.1}_{-0.1}\times 10^{-4}$ & 0.93 (1025) & 42.4 \\
\noalign{\smallskip}
IRAS F05189-2524 & 60201022002 & 155 & 24.7 & $1.57 \pm 0.09$ & $8.5^{+0.2}_{-0.2}\times 10^{-5}$ & 0.97 (761) & 43.3 \\
\noalign{\smallskip}
ESO 253-G003 & 60802002024 & 26 & 23.6 & $1.50 \pm 0.05$ & $3.0^{+0.3}_{-0.2}\times 10^{-5}$ & 0.98 (98) & 43.1 \\
\noalign{\smallskip}
Mrk 6 & 60102044004 & 62 & 23.5 & $1.93 \pm 0.01$ & $3.3^{+0.1}_{-0.1}\times 10^{-4}$ & 1.03 (793) & 43.1 \\
\noalign{\smallskip}
Mrk 9 & 60061326002 & 23 & 22.0 & $1.88 \pm 0.03$ & $4.4^{+0.3}_{-0.3}\times 10^{-5}$ & 1.06 (174) & 43.4 \\
\noalign{\smallskip}
Mrk 79 & 60601010002 & 66 & 22.2 & $2.03 \pm 0.01$ & $03.2^{+0.1}_{-0.1}\times 10^{-4}$ & 0.97 (980) & 43.8 \\
\noalign{\smallskip}
Mrk 704 & 60061090002 & 22 & 23.0 & $1.91 \pm 0.01$ & $1.7^{+0.1}_{-0.1}\times 10^{-4}$ & 1.05 (387) & 43.4 \\
\noalign{\smallskip}
NGC 2992 & 90501623002 & 21 & 22.1 & $1.77 \pm 0.01$ & $5.8^{+0.1}_{-0.1}\times 10^{-4}$ & 0.98 (1505) & 43.2 \\
\noalign{\smallskip}
Mrk 1239 & 60360006002 & 21 & 23.0 & $1.88 \pm 0.04$ & $5.0^{+0.3}_{-0.3}\times 10^{-5}$ & 0.98 (150) & 43.3 \\
\noalign{\smallskip}
3C 234 & 60160382002 & 15 & 23.3 & $1.64 \pm 0.05$ & $2.0^{+0.2}_{-0.2}\times 10^{-5}$ & 0.80 (99) & 44.5 \\
\noalign{\smallskip}
NGC 3079 & 60061097002 & 22 & 24.5 & $1.91 \pm 0.06$ & $1.1^{+0.2}_{-0.2}\times 10^{-3}$ & 1.07 (75) & 42.1 \\
\noalign{\smallskip}
NGC 3227 & 60202002004 & 42 & 22.2 & $1.80 \pm 0.01$ & $3.4^{+0.1}_{-0.1}\times 10^{-4}$ & 0.99 (1017) & 42.0 \\
\noalign{\smallskip}
NGC 3516 & 60160001002 & 33 & 22.4 & $1.93 \pm 0.01$ & $4.7^{+0.1}_{-0.1}\times 10^{-4}$ & 1.01 (1045) & 42.7 \\
\noalign{\smallskip}
NGC 3982 & 60375001002 & 33 & 24.0 & $1.92 \pm 0.20$ & $2.7^{+1.3}_{-0.9}\times 10^{-4}$ & 0.91 (24) & 40.2 \\
\noalign{\smallskip}
NGC 4051 & 60001050002 & 9 & 22.6 & $2.08 \pm 0.02$ & $4.8^{+0.1}_{-0.1}\times 10^{-4}$ & 1.04 (397) & 41.6 \\
\noalign{\smallskip}
UGC 7064 & 60601035002 & 77 & 22.5 & $1.74 \pm 0.02$ & $6.5^{+0.2}_{-0.2}\times 10^{-5}$ & 0.92 (525) & 42.5 \\
\noalign{\smallskip}
NGC 4151 & 60502017006 & 32 & 23.0 & $1.83 \pm 0.01$ & $2.4^{+0.1}_{-0.1}\times 10^{-3}$ & 1.10 (1822) & 43.0 \\
\noalign{\smallskip}
Mrk 766 & 60101022002 & 24 & 22.3 & $2.24 \pm 0.01$ & $4.0^{+0.1}_{-0.1}\times 10^{-4}$ & 0.99 (559) & 43.2 \\
\noalign{\smallskip}
NGC 4388 & 60061228002 & 21 & 23.6 & $1.66 \pm 0.04$ & $3.3^{+0.5}_{-0.4}\times 10^{-4}$ & 1.04 (420) & 42.9 \\
\noalign{\smallskip}
NGC 4593 & 60001149002 & 23 & 20.3 & $1.88 \pm 0.01$ & $2.3^{+0.1}_{-0.1}\times 10^{-4}$ & 1.01 (618) & 43.1 \\
\noalign{\smallskip}
NGC 4748 & 60663002002 & 81 & 22.2 & $2.19 \pm 0.01$ & $1.3^{+0.1}_{-0.1}\times 10^{-4}$ & 1.01 (623) & 42.5 \\
\noalign{\smallskip}
Mrk 231 & 90102001006 & 31 & 22.8 & $1.71 \pm 0.07$ & $9.3^{+1.1}_{-1.1}\times 10^{-6}$ & 0.94 (67) & 42.5 \\
\noalign{\smallskip}
NGC 4941 & 60061236002 & 21 & 23.6 & $1.62 \pm 0.05$ & $5.6^{+0.6}_{-0.5}\times 10^{-5}$ & 0.91 (79) & 41.3 \\
\noalign{\smallskip}
NGC 5033 & 60601023002 & 104 & 22.1 & $1.92 \pm 0.01$ & $1.2^{+0.1}_{-0.1}\times 10^{-4}$ & 1.01 (818) & 41.1 \\
\noalign{\smallskip}
MCG -03-34-064 & 60101020002 & 79 & 23.7 & $2.03 \pm 0.01$ & $4.6^{+0.1}_{-0.1}\times 10^{-4}$ & 1.01 (717) & 42.9 \\
\noalign{\smallskip}
NGC 5135 & 60001153002 & 33 & 23.3 & $1.43 \pm 0.06$ & $2.0^{+0.1}_{-0.1}\times 10^{-5}$ & 0.90 (80) & 43.3 \\
\noalign{\smallskip}
NGC 5194 & 60201062003 & 163 & 24.4 & $1.91 \pm 0.03$ & $3.6^{+0.2}_{-0.2}\times 10^{-5}$ & 0.99 (224) & 41.2 \\
\noalign{\smallskip}
MCG -06-30-015 & 60902004002 & 131 & 22.1 & $2.13 \pm 0.01$ & $5.7^{+0.1}_{-0.1}\times 10^{-4}$ & 1.10 (1386) & 42.9 \\
\noalign{\smallskip}
IRAS F13349+2438 & 60501019002 & 50 & 23.3 & $2.72 \pm 0.14$ & $4.0^{+0.1}_{-0.5}\times 10^{-4}$ & 1.09 (175) & 43.8 \\
\noalign{\smallskip}
Mrk 266 & 60465005002 & 32 & 24.3 & $1.93 \pm 0.15$ & $1.5^{+0.5}_{-0.5}\times 10^{-5}$ & 0.89 (42) & 42.3 \\
\noalign{\smallskip}
Mrk 273 & 60002028002 & 70 & 23.5 & $1.78 \pm 0.02$ & $3.7^{+0.1}_{-0.1}\times 10^{-5}$ & 0.89 (249) & 43.0 \\
\noalign{\smallskip}
IC 4329A & 60701015002 & 82 & 22.0 & $1.84 \pm 0.01$ & $7.5^{+0.1}_{-0.1}\times 10^{-4}$ & 1.01 (1582) & 43.7 \\
\noalign{\smallskip}
NGC 5347 & 60001163002 & 47 & 24.6 & $2.26 \pm 0.10$ & $1.3^{+0.3}_{-0.3}\times 10^{-4}$ & 1.20 (64) & 41.3 \\
\noalign{\smallskip}
Mrk 463 & 60061249002 & 24 & 23.5 & $1.73 \pm 0.05$ & $3.1^{+0.2}_{-0.2}\times 10^{-5}$ & 0.88 (62) & 43.2 \\
\noalign{\smallskip}
NGC 5506 & 60501015002 & 61 & 22.7 & $1.86 \pm 0.01$ & $8.0^{+0.1}_{-0.1}\times 10^{-4}$ & 0.97 (1410) & 42.8 \\
\noalign{\smallskip}
NGC 5548 & 60002044006 & 51 & 21.7 & $1.85 \pm 0.01$ & $3.8^{+0.1}_{-0.1}\times 10^{-4}$ & 1.05 (1099) & 43.4 \\
\noalign{\smallskip}
Mrk 817 & 60601007002 & 135 & 22.8 & $1.97 \pm 0.01$ & $1.7^{+0.1}_{-0.1}\times 10^{-4}$ & 1.03 (1004) & 43.4 \\
\noalign{\smallskip}
IRAS F15091-2107 & 60061262002 & 21 & 22.6 & $2.02 \pm 0.02$ & $1.6^{+0.1}_{-0.1}\times 10^{-4}$ & 1.04 (403) & 43.8 \\
\noalign{\smallskip}
NGC 5929 & 60465009002 & 29 & 23.0 & $1.77 \pm 0.05$ & $3.1^{+0.1}_{-0.1}\times 10^{-5}$ & 0.74 (134) & 41.5 \\
\noalign{\smallskip}
MCG -02-40-004 & 60061267002 & 21 & 24.3 & $1.91 \pm 0.02$ & $2.9^{+0.1}_{-0.1}\times 10^{-4}$ & 0.99 (507) & 43.3 \\
\noalign{\smallskip}
IRAS F15480-0344 & 90601603002 & 48 & 24.8 & $1.69 \pm 0.05$ & $4.0^{+0.4}_{-0.4}\times 10^{-5}$ & 0.80 (89) & 42.9 \\
\noalign{\smallskip}
ESO 141-G055 & 60201042002 & 93 & 21.5 & $1.96 \pm 0.01$ & $2.2^{+0.1}_{-0.1}\times 10^{-4}$ & 1.00 (1153) & 43.9 \\
\noalign{\smallskip}
NGC 6890 & 60375003002 & 35 & 23.3 & $2.08 \pm 0.03$ & $1.7^{+0.1}_{-0.1}\times 10^{-4}$ & 1.02 (252) & 41.9 \\
\noalign{\smallskip}
Mrk 509 & 60101043004 & 36 & 20.0 & $1.77 \pm 0.01$ & $3.9^{+0.1}_{-0.1}\times 10^{-4}$ & 1.03 (918) & 44.0 \\
\noalign{\smallskip}
IC 5063 & 60061302002 & 18 & 23.3 & $1.78 \pm 0.01$ & $3.4^{+0.1}_{-0.1}\times 10^{-4}$ & 1.00 (513) & 42.6 \\
\noalign{\smallskip}
Mrk 897 & 60701042002 & 21 & 23.7 & $1.95 \pm 0.07$ & $5.7^{+0.9}_{-0.8}\times 10^{-5}$ & 0.85 (65) & 42.9 \\
\noalign{\smallskip}
NGC 7172 & 60061308002 & 32 & 23.0 & $1.90 \pm 0.02$ & $6.7^{+0.9}_{-0.6}\times 10^{-4}$ & 1.00 (1150) & 43.0 \\
\noalign{\smallskip}
3C 445 & 60160788002 & 20 & 23.3 & $1.66 \pm 0.04$ & $2.5^{+1.2}_{-0.5}\times 10^{-4}$ & 1.04 (310) & 44.0 \\
\noalign{\smallskip}
NGC 7314 & 60201031002 & 100 & 22.1 & $2.03 \pm 0.01$ & $4.9^{+0.1}_{-0.1}\times 10^{-4}$ & 1.03 (1276) & 43.1 \\
\noalign{\smallskip}
NGC 7469 & 60101001008 & 23 & 21.8 & $1.99 \pm 0.01$ & $2.8^{+0.1}_{-0.1}\times 10^{-4}$ & 1.02 (681) & 43.2 \\
\noalign{\smallskip}
NGC 7582 & 60201003002 & 48 & 23.5 & $1.78 \pm 0.01$ & $4.0^{+0.1}_{-0.1}\times 10^{-4}$ & 1.06 (1028) & 42.4 \\
%\noalign{\smallskip}
%NGC 7590 & 60201003002 & 48 & 23.5 & $1.78 \pm 0.01$ & $4.0^{+0.1}_{-0.1}\times 10^{-4}$ & 1.04 (1092) & 42.4 \\
\noalign{\smallskip}
NGC 7674 & 60001151002 & 52 & 23.3 & $1.63 \pm 0.03$ & $2.8^{+0.1}_{-0.1}\times 10^{-5}$ & 1.01 (200) & 42.6 \\
\end{longtable}
\flushleft
Columns: 1 = source name, 2 = \nustar\ observation ID, 3 = effective exposure time, 4 = line-of-sight column density, 5 = photon index, 6 = normalization of the BMC model in units of $\mathrm{10^{39}\,erg\,s^{-1}\,kpc^{-2}}$, 7 = reduced chi-square and degrees of freedom, 8 = X-ray luminosity in the 2--10 keV energy band.
} % end \longtab

\longtab[3]{
\begin{longtable}{lcccccll}
\caption{Black hole masses} \\
\label{tab:BH_masses} \\

\toprule
\midrule
Source & Distance  &$\log(\lambda_\mathrm{Edd})$ & $\log \left(\frac{M_\textrm{BH,X}}{M_\odot}\right)$ & $\log \left(\frac{M_\textrm{BH,dyn}}{M_\odot}\right)$ & $\Delta \log(M_\mathrm{BH})/\sigma_\mathrm{diff}$ & Method & Reference \\
     &   [Mpc]    &           &      &  & & &\\
(1) & (2) & (3) & (4) & (5) & (6) & (7) & (8) \\
\midrule
\endfirsthead

\caption{continued.} \\
\toprule
\midrule
Source & Distance  &$\log(\lambda_\mathrm{Edd})$ & $\log \left(\frac{M_\textrm{BH,X}}{M_\odot}\right)$ & $\log \left(\frac{M_\textrm{BH,dyn}}{M_\odot}\right)$ & $\Delta\log(M_\mathrm{BH})/\sigma_\mathrm{diff}$ & Method & Reference \\
     &   [Mpc]    &           &      &  & & & \\
(1) & (2) & (3) & (4) & (5) & (6) & (7) & (8) \\
\midrule
\endhead

\bottomrule
\endfoot

\bottomrule
\endlastfoot
Mrk 335 & 107 & -0.41 & $7.16 \pm 0.12$ & $7.25 \pm 0.11$ & 0.6 & RM model& \citet{Villafana2026} \\
Mrk 348 & 60.6 & -1.79 & $8.19 \pm 0.16$ & $8.09 \pm 0.60$& 0.2 & SE & \citet{LaMassa2025} \\
UGC 545 & 269 & 0.09 & $7.46 \pm 0.15$ & $7.26 \pm 0.60$  & 0.3 & SE & \citet{Vestergaard2002} \\
NGC 424 & 43.3 & -0.56 & $6.85 \pm 0.28$ & $\dots$ & $\dots$ & $\dots$ & $\dots$ \\
NGC 513 & 80.8 & -1.25 & $7.04 \pm 0.21$ & $\dots$ & $\dots$ & $\dots$ & $\dots$ \\
IRAS F01475-0740 & 72.7 & 0.18 & $6.13 \pm 0.28$ & $\dots$   & $\dots$  & $\dots$ & $\dots$ \\
NGC 931 & 68.8 & -1.00 & $7.38 \pm 0.06$ & $6.84 \pm 0.40$ & 1.3 & RM & \citet{Zastrocky2024} \\
NGC 1068 & 10.1 & -0.26 & $6.60 \pm 0.05$ & $6.75 \pm 0.02$ & 2.8 & maser & \citet{Lodato2003} \\
NGC 1125 & 44.5 & -0.99 & $6.51 \pm 0.26$ & $7.24 \pm 0.60$ & 1.1 & SE & \citet{LaMassa2025} \\
NGC 1144 & 128 & -0.75 & $7.61 \pm 0.15$ & $\dots$ & $\dots$ & $\dots$ & $\dots$ \\
MCG -02-08-039 & 128 & $\dots$ & $6.99\pm 0.30$ & $\dots$ & $\dots$ & $\dots$ & $\dots$\\
NGC 1194 & 56.2 & -1.57 & $6.80 \pm 0.06$ & $7.84 \pm 0.02$ &16.4 & maser & \citet{Kuo2011} \\
NGC 1320 & 37.7 & -0.28 & $5.98 \pm 0.17$ & $6.63 \pm 0.21$ & 2.4 & maser & \citet{Greene2016} \\
NGC 1365 & 18.1 & -0.50 & $6.16 \pm 0.22$ & $6.65 \pm 0.60$ & 0.8 & SE & \citet{Onori2017} \\
NGC 1386 & 15.9 & 0.03 & $5.40 \pm 0.38$ & $6.07 \pm 0.29$ & 1.4 & maser & \citet{Braatz1997} \\
IRAS F03362-1642 & 161 & -1.54 & $7.20 \pm 0.21$ & $\dots$ & $\dots$ & $\dots$ & $\dots$\\
NGC 1566 & 17.9 & -1.27 & $5.91 \pm 0.21$ & $6.29 \pm 0.60$ & 0.6 & SE & \citet{Ho2008} \\
3C 120 & 127 & -1.08 & $8.36 \pm 0.07$ & $7.84 \pm 0.18$ & 2.7 & RM model & \citet{Villafana2026} \\
Mrk 618 & 156 & -1.16 & $7.80 \pm 0.17$ & $7.17 \pm 0.40$ & 1.4 & RM & \citet{Zastrocky2024} \\
IRAS F04385-0828 & 64.8 & $\dots$ & $6.61 \pm 0.22$ & $\dots$ & $\dots$ & $\dots$ & $\dots$\\
ESO 033-G002 & 84 & -1.43 & $7.43 \pm 0.20$ & $7.15 \pm 0.60$ & 0.4 & SE & \citet{LaMassa2025} \\
ESO 362-G018 & 54.3 & -1.16 & $6.95 \pm 0.19$ & $7.19 \pm 0.60$ &0.4 &  SE & \citet{Ho2008} \\
IRAS F05189-2524 & 196 & -0.35 & $7.65 \pm 0.10$ & $\dots$ & $\dots$ & $\dots$ & $\dots$\\
ESO 253-G003 & 189 & $\dots$ & $7.73 \pm 0.17$ & $\dots$ & $\dots$ & $\dots$ & $\dots$\\
Mrk 6 & 84.8 & -1.20 & $7.60 \pm 0.29$ & $8.10 \pm 0.40$ & 1.0 & RM & \citet{Bentz2015} \\
Mrk 9 & 174 & -0.86 & $7.41 \pm 0.21$ & $7.61 \pm 0.60$ & 0.3 & SE & \citet{U2022} \\
Mrk 79 & 98.4 & -0.89 & $7.68 \pm 0.07$ & $7.61 \pm 0.40$ & 0.2 & RM & \citet{Bentz2015} \\
Mrk 704 & 130 & -0.95 & $7.72 \pm 0.17$ & $7.57 \pm 0.40$ & 0.4 & RM & \citet{Bentz2015} \\
NGC 2992 & 29.8 & -1.15 & $7.11 \pm 0.16$ & $7.36 \pm 0.60$ & 0.4 & SE & \citet{Guolo2021} \\
Mrk 1239 & 91.8 & -0.41 & $6.91 \pm 0.22$ & $7.47 \pm 0.50$ & 1.0 & RM model & \citet{Gravity2023} \\
3C 234 & 906 & -0.68 & $8.68 \pm 0.18$ & $\dots$ & $\dots$ & $\dots$ & $\dots$\\
NGC 3079 & 20.6 & -1.62 & $6.90 \pm 0.13$ & $6.36 \pm 0.09$ & 3.4 & maser & \citet{Yamauchi2004} \\
NGC 3227 & 23.0 & -1.44 & $6.62 \pm 0.15$ & $7.04 \pm 0.11$ & 2.3 & RM model & \citet{Bentz2023} \\
NGC 3516 & 38 & -1.12 & $7.05 \pm 0.26 $ & $7.40 \pm 0.40$ & 0.7 & RM & \citet{Bentz2015} \\
NGC 3982 & 22.3 & -0.66 & $6.44 \pm 0.36$ & $< 6.84$ & $\dots$ & gas dynamics & \citet{Beifiori2012} \\
NGC 4051 & 17.6 & -1.08 & $5.79 \pm 0.12$ & $5.89 \pm 0.40$ & 0.2 & RM & \citet{Bentz2015} \\
UGC 7064 & 113 & -1.59 & $7.36 \pm 0.22$ & $7.5 \pm 0.60$ & 0.2 & SE & \citet{LaMassa2025} \\
NGC 4151 & 15.8 & -1.32 & $7.11 \pm 0.17$ & $7.23 \pm 0.10$ & 0.6 & RM model & \citet{Bentz2022} \\
Mrk 766 & 59.9 & -0.99 & $7.14 \pm 0.16$ & $6.82 \pm 0.40$ & 0.7 & RM & \citet{Bentz2015} \\
NGC 4388 & 18 & -0.65 & $6.67 \pm 0.22$ & $6.91 \pm 0.01$ & 1.1 & maser & \citet{Kuo2011} \\
NGC 4593 & 37.2 & -0.69 & $6.78 \pm 0.13$ & $6.88 \pm 0.40$ & 0.2 & RM & \citet{Bentz2015} \\
NGC 4748 & 68.3 & -1.10 & $6.80 \pm 0.16$ & $6.41 \pm 0.40$ & 0.9 & RM & \citet{Bentz2015} \\
Mrk 231 & 189 & 0.47 & $7.00 \pm 0.30$ & $\dots$ & $\dots$ & $\dots$ & $\dots$\\
NGC 4941 & 21.2 & -1.14 & $5.89 \pm 0.18$ & $\dots$ & $\dots$ & $\dots$ & $\dots$\\
NGC 5033 & 19.1 & -1.83 & $5.88 \pm 0.18$ & $\dots$ & $\dots$ & $\dots$ & $\dots$\\
MCG -03-34-064 & 79.1 & -1.22 & $7.65 \pm 0.08$ & $\dots$ & $\dots$ & $\dots$ & $\dots$\\
NGC 5135 & 63.5 & -0.67 & $6.96 \pm 0.36$ & $\dots$ & $\dots$ & $\dots$ & $\dots$\\
NGC 5194 & 7.6 & -0.90 & $4.64 \pm 0.12$ & $< 5.58$ & $\dots$ & gas dynamics & \citet{Beifiori2012} \\
MCG -06-30-015 & 37.3 & -1.08 & $6.97 \pm 0.15$ & $6.20 \pm 0.40$ & 1.8 & RM & \citet{Bentz2016} \\
IRAS F13349+2438 & 503 & -0.50 & $8.52 \pm 0.27$ & $\dots$ & $\dots$ & $\dots$ & $\dots$\\
Mrk 266 & 124 & -0.81 & $6.67 \pm 0.30$ & $\dots$ & $\dots$ & $\dots$ & $\dots$\\
Mrk 273 & 166 & -0.90 & $7.40 \pm 0.20$ & $\dots$ & $\dots$ & $\dots$ & $\dots$\\
IC 4329A & 73.6 & -0.86 & $7.94 \pm 0.15$ & $7.63 \pm 0.40$ & 0.7 & RM model & \citet{Villafana2026} \\
NGC 5347 & 21.8 & -0.44 & $5.75 \pm 0.49$ & $< 7.21$ & $\dots$ & gas dynamics & \citet{Beifiori2012} \\
Mrk 463 & 230 & -0.35 & $7.66 \pm 0.42$ & $\dots$ & $\dots$ & $\dots$ & $\dots$\\
NGC 5506 & 23.8 & -0.89 & $6.96 \pm 0.14$ & $\dots$ & $\dots$ & $\dots$ & $\dots$\\
NGC 5548 & 77.6 & -1.25 & $7.68 \pm 0.16$ & $7.51 \pm 0.16$ & 0.8 & RM model & \citet{Villafana2026} \\
Mrk 817 & 139 & -0.94 & $7.78 \pm 0.07$ & $7.59 \pm 0.40$ & 0.5 & RM & \citet{Bentz2015} \\
IRAS F15091-2107 & 201 & -0.91 & $8.03 \pm 0.09$ & $7.00 \pm 0.60$ & 1.7 & SE & \citet{Ohta2007} \\
NGC 5929 & 38.5 & -1.10 & $6.07 \pm 0.27$ & $\dots$ & $\dots$ & $\dots$ & $\dots$\\
MCG -02-40-004 & 112 & -1.13 & $7.80 \pm 0.20$ & $\dots$ & $\dots$ & $\dots$ & $\dots$\\
IRAS F15480-0344 & 135 & -0.76 & $7.27 \pm 0.17$ & $\dots$ & $\dots$ & $\dots$ & $\dots$\\
ESO 141-G055 & 163 & -0.84 & $8.04 \pm 0.07$ & $\dots$ & $\dots$ & $\dots$ & $\dots$\\
NGC 6890 & 31.8 & -0.85 & $6.34 \pm 0.16$ & $\dots$ & $\dots$ & $\dots$ & $\dots$\\
Mrk 509 & 147 & -0.70 & $8.33 \pm 0.18$ & $8.05 \pm 0.40$ & 0.6 & RM & \citet{Bentz2015} \\
IC 5063 & 37.5 & -0.80 & $7.05 \pm 0.18$ & $7.74 \pm 0.60$ & 1.1 & SE & \citet{LaMassa2025} \\
Mrk 897 & 110 & -1.20 & $7.13 \pm 0.21$ & $\dots$ & $\dots$ & $\dots$ & $\dots$\\
NGC 7172 & 33.9 & -1.24 & $7.15 \pm 0.19$ & $\dots$ & $\dots$ & $\dots$ & $\dots$\\
3C 445 & 244 & -1.35 & $8.60 \pm 0.38$ & $\dots$ & $\dots$ & $\dots$ & $\dots$\\
NGC 7314 & 16 & -0.85 & $6.18 \pm 0.05$ & $6.41 \pm 0.60$ & 0.4 & SE & \citet{LaMassa2025} \\
NGC 7469 & 65.1 & -0.71 & $7.32 \pm 0.07$ & $6.96 \pm 0.40$ & 0.9 & RM & \citet{Bentz2015} \\
NGC 7582 & 22.5 & -1.99 & $6.70 \pm 0.16$ & $7.74 \pm 0.09$ & 5.7 & gas dynamics & \citet{Wold2006} \\
%NGC 7590 & 31.1 & -2.32 & $6.99 \pm 0.18$ & $\dots$ & $\dots$ & $\dots$ & $\dots$\\
NGC 7674 & 122 & -0.24 & $7.10 \pm 0.12$ & $\dots$ & $\dots$ & $\dots$ & $\dots$\\
\end{longtable}
\flushleft
Columns: 1 = source name, 2 = distance, 3 = Eddington ratio \laedd=$L_\mathrm{bol}/L_\mathrm{Edd}$, 4 = \mbh\ determined with the X-ray method, 5 = \mbh\ determined with dynamical methods, 6 = statistics used to quantify the difference between X-ray scaling and dynamical estimates of log(\mbh) in terms of sigmas,   7 = dynamical method used to determine the \mbh\ reported in column 5 (RM model = reverberation mapping with BLR modeling, RM = standard reverberation mapping, SE = single-epoch method), 8 = reference of the dynamical method estimates. Columns 5--8 are filled in only for those \mbh\ values determined with dynamical methods.
} % end \longtab

\end{appendix}
\end{document}